\documentclass[final]{iopart}
\usepackage{iopams,graphicx,texdraw,epsf,bm,texdraw}
\begin{document}
\paper[Large-$n$ expansion\ldots]{Large-$\boldsymbol{n}$ expansion
  for $\boldsymbol{m}$-axial Lifshitz points$^\heartsuit$}

\author[M A Shpot \etal ]{M A Shpot{\dag}{\ddag}, Yu M Pis'mak{\S}{\ddag}
  and H W Diehl{\ddag}}
\address{{\dag}\ Institute for Condensed Matter Physics, 79011 Lviv, Ukraine}
\address{{\ddag}\ Fachbereich Physik, Universit{\"a}t
Duisburg-Essen, Campus Essen, D-45117 Essen, Federal Republic of Germany}
\address{{\S}\ State University of Sankt-Petersburg, 198504
  Sankt-Petersburg, Russia}

\ead{shpot@ph.icmp.lviv.ua}
\setcounter{footnote}{3}
\begin{abstract}
  The large-$n$ expansion is developed for the study of critical
  behaviour of $d$-dimensional systems at $m$-axial Lifshitz points
  with an arbitrary number $m$ of modulation axes. The leading
  non-trivial contributions of $\Or(1/n)$ are derived for the two
  independent correlation exponents $\eta_{L2}$ and $\eta_{L4}$, and the
  related anisotropy index $\theta$. The series coefficients of these
  $1/n$ corrections are given for general values of $m$ and $d$ with
  $0 \leq m\leq d$ and $2+m/2<d<4+m/2$ in the form of integrals. For
  special values of $m$ and $d$ such as $(m,d)=(1,4)$, they can be
  computed analytically, but in general their evaluation requires
  numerical means. The $1/n$ corrections are shown to reduce in the
  appropriate limits to those of known large-$n$ expansions for the
  case of $d$-dimensional isotropic Lifshitz points and critical
  points, respectively, and to be in conformity with available
  dimensionality expansions about the upper and lower critical
  dimensions.  Numerical results for the $1/n$ coefficients of
  $\eta_{L2}$, $\eta_{L4}$ and $\theta$ are presented for the physically
  interesting case of a uniaxial Lifshitz point in three dimensions,
  as well as for some other choices of $m$ and $d$. A universal
  coefficient associated with the energy-density pair correlation
  function is calculated to leading order in $1/n$ for general values
  of $m$ and $d$.

\end{abstract}
\submitto{\JPCM} \vspace{10pt }
     \noindent{\rm $^\heartsuit$Dedicated to: {\it Lothar Sch{\"a}fer
         on the occasion of his 60th birthday} \par}

\section{Introduction}\label{sec:intro}

Systems exhibiting critical behaviour can be divided into
universality classes such that all members of a given class have the
same universal critical properties (Fisher 1983). The universality
classes are represented by field theories, such as the $n$-component
$\phi^4$ models in $d$ space dimensions, which are nontrivial whenever
$d$ is less than the upper critical dimension $d^*$ above which Landau
theory holds. Proper analyses of such field theories usually require
sophisticated tools such as renormalization group approaches,
elaborate numerical simulations or a combination of both (Wilson and
Kogut 1974, Domb and Green 1976).

The commonly employed successful analytical methods are dimensionality
expansions about the upper and lower critical dimensions $d^*$ and
$d_\ell$ (Wilson and Fisher 1972, Polyakov 1975, Br{\'e}zin and
Zinn-Justin 1976a, 1976b, Bardeen \etal 1976), the massive
renormalization group (RG) approach in fixed dimensions (Parisi 1980),
and the $1/n$ expansion in inverse powers of the number $n$ of
order-parameter components (Ma 1973, Abe 1973). Remarkably precise
estimates of critical indices and other universal quantities of
three-dimensional systems have been obtained both via $\epsilon=d^*-d$
expansions and the massive RG approach at fixed $d$ by computing
perturbation series to sufficiently high orders and resumming them
(Guida and Zinn-Justin 1998, Pelissetto and Vicari 2002, Privman \etal
1991).  Expansions about the lower critical dimension $d_\ell$, on the
other hand, seem to have a more modest potential for precise
estimates, unless they are combined with information from other
sources.

In this paper we will be concerned with the $1/n$~expansion. Our aim
is to develop this approach for the study of critical behaviour at
$m$-axial Lifshitz points. A Lifshitz point (LP) is a multicritical
point at which a disordered phase meets both a spatially uniform
ordered and a spatially modulated ordered phase (Hornreich \etal
1975a, Hornreich 1980, Selke 1992, Diehl 2002, 2004). The disordered
phase is separated from the two ordered ones by a line of critical
temperatures $T_c=T_c(g)$, depending on a thermodynamic nonordering
field $g$, such as pressure or a ratio of an antiferromagnetic and a
ferromagnetic coupling. The LP, located at $T_L=T_c(g_L)$, divides
this critical line into two sections. In the modulated ordered phase,
the wave-vector associated with the modulation, $\bm{q}_{\rm
  mod}=\bm{q}_{\rm mod}(T,g)$, varies with $g$ and temperature $T$.
Without loss of generality, we can consider the ferromagnetic case.
Then uniform order corresponds to $\bm{q}_{\rm mod}=\bm{0}$, and
$\bm{q}_{\rm mod}(T,g)$ vanishes at the {LP}. The LP is called
$m$-axial if the wave-vector instability that sets in at the LP occurs
in an $m$-dimensional subspace of $d$-dimensional space, i.e.\
$\bm{q}_{\rm mod}\in\mathbb{R}^m$ with $0\leq m \leq d$. The limiting
values $m=0$ and $m=d$ correspond to the cases of a usual critical
point (CP) and the isotropic LP, respectively.

As is well known, the large-$n$ expansion leads to self-consistent
equations (Abe 1973, Br{\'e}zin \etal 1976, Vasiliev \etal 1981a).
These are considerably more difficult to handle than
ordinary perturbation expansions, which is the essential input
required in the alternative approaches mentioned above (dimensionality
expansions, massive RG approach). For this reason the available series
expansions in $1/n$ are restricted to low orders, even for the much
simpler case of a conventional {CP}. Furthermore, $1/n$
expansions normally converge slowly.%
\footnote{According to some recent results (Baym \etal 2000, Arnold
  and Tom{\'a}{\v{s}}ik 2000), the shift of the critical temperature of a
  dilute Bose gas seems to be an exception
  to this rule.} %
On the other hand, this technique has a number of very attractive
features.  One is its capability of treating fluctuation effects in a
systematic, nonperturbative manner.%
\footnote{A familiar numerical RG scheme having this capablility
  employs the so-called ``exact'' RG equations (Wilson and Kogut 1974).
  Its only application to critical behaviour at LPs we are aware of is
  the recent work by Bervillier (2004). This deals with the case of a
  uniaxial LP with $n=1$ in $d=3$ dimensions and uses the lowest
  ("local potential") approximation.}  Another is that it can be
applied for any fixed value of $d$ between $d_\ell$ and $d^*$.  No
additional expansion in a small parameter such as $\epsilon=d^*-d$ is
required.
%
%Similar features have also the so-called ``exact'' RG equations
%(Wilson and Kogut 1974).  Except for the recent work w the lowest
%("local potential") approximation by uniaxial LP with $n=1$ in three
%dimensions we are not aware of any studies of critical b have been
%recently been applied to the study of by in the case of.
%
Owing to these appealing properties the $1/n$ expansion continues to
be an important tool for the analysis of nontrivial field theories.
The spectrum of problems to which it has recently been applied and
made significant contributions is impressively rich. It ranges from
the study of classical spin models (Br{\'e}zin 1993, Vasiliev 1998,
Zinn-Justin 1996, Moshe and Zinn-Justin 2003, Campostrini \etal 1998,
Pelissetto \etal 2001, Gracey 2002a, 2002b), critical behaviour in
bounded systems (McAvity and Osborn 1995), the physics of disordered
elastic media (Le Doussal and Wiese 2003, 2004) and models of
stochastic surface growth (Doherty \etal 1994) to problems of
high-energy physics (Aharony \etal 2000, Moshe and Zinn-Justin 2003)
and quantum phase transitions (Franz \etal 2002).

Unfortunately, for critical behaviour at LPs, hardly any results have
yet been worked out by means of it. %this expansion.
Previous large-$n$ analyses for general values of $m\in [0,d]$ (Kalok
and Obermair 1976, Hornreich \etal 1977, Mukamel and Luban 1978,
Frachebourg and Henkel 1993) were restricted to the limit $n\to\infty$. To
our knowledge, the only exceptions in which corrections to order $1/n$
were computed deal exclusively with the case $m=d$ of a $d$-axial LP
(Hornreich \etal 1975b, Nicoll \etal 1976, Inayat-Hussain and
Buckingham 1990).

This lack of results for general $m$ is due to the complications
arising from the combination of two special features such systems
have: (i) the \emph{anisotropic} nature of scale invariance they
exhibit at LPs, and (ii) the \emph{unusually complicated form} of the
two-point scaling functions they have in position space already at the
level of Landau theory (Diehl and Shpot 2000, Shpot and Diehl 2001).
Anisotropic scale invariance means that coordinate separations
$dz_\alpha$ along some directions scale as nontrivial powers of their
counterparts $dr_\beta$ along the remaining orthogonal ones, i.e.\
$dz_\alpha\sim (dr_\beta)^\theta$, where $\theta$, called anisotropy
exponent, differs from 1. Of course, anisotropic scale invariance is a
feature encountered also in studies of dynamical critical behaviour at
usual CPs, where time $t$ scales as a nontrivial power of the spatial
separations, and for which analyses to $\Or(1/n)$ can be found in the
literature (see e.g.\ Halperin \etal (1972); for more recent uses of
the $1/n$ expansion in dynamics, see e.g.\ Bray (2002) or Moshe and
Zinn-Justin (2003)). It is the combination with (ii) that makes
consistent treatments of fluctuation effects so hard for critical
behaviour at $m$-axial {LPs}.  The very same difficulties had
prevented the performance of full two-loop RG analyses within the
framework of the $\epsilon=d^*-d$ expansion for decades, and produced
controversial $\Or(\epsilon^2)$ results (Mukamel 1977, Sak and Grest
1978, Hornreich and Bruce 1978, Hornreich 1980, Mergulh{\~a}o and
Carneiro 1999) until such a RG analysis for general $m$ was finally
accomplished in (Diehl and Shpot 2000, Shpot and Diehl
2001).%
\footnote{For a discussion and critical assessment of recent work
  (de Albuquerque and Leite 2001, Leite 2003) giving alternative results,
  see (Diehl and Shpot 2001, 2003).}

In order to overcome the mentioned technical difficulties, we shall
adapt and extend the elegant technique suggested by Vasiliev \etal\
(1981a) and utilized in subsequent work (Vasiliev \etal 1981b) to
compute the $1/n$ expansion of the standard critical exponents $\eta$
and $\nu$ for the case $m=0$ of a CP to $\Or(1/n^{2})$. (In Vasiliev
\etal (1981c), the authors managed to compute $\eta$ even to
$\Or(1/n^3)$ by means of a conformal bootstrap method.) We proceed as
follows:

We start in section~\ref{sec:scalforms} by specifying the expected
scaling forms of the required two-point functions.  These are employed
in section~\ref{sec:sceq} to solve the resulting self-consistent
equations to the appropriate order $1/n$. Matching the
anticipated asymptotic large-distance behaviour of the cumulants
with the solutions of these equations yields consistency conditions.
From these, the $\Or(1/n)$ terms of the two independent correlation
exponents $\eta_{L2}$ and $\eta_{L4}$ can be determined for general values
of $m$ in the form of finite, numerically computable integrals.

In section~\ref{sec:checkE} we verify that these coefficients reduce
in the limits $m\to 0$ and $m\to d$ to the analytical expressions for
the CP case and the isotropic LP obtained in (Ma 1973, Abe and Hikami
1973) and (Hornreich \etal 1975b), respectively. In
section~\ref{sec:epsexp} we consider the behaviour of our $\Or(1/n)$
results as $\epsilon=4+m/2-d\to 0$, and explicitly reproduce the large-$n$
limits of the $\Or(\epsilon^2)$ coefficients of (Diehl and Shpot 2000, Shpot
and Diehl 2001) for the correlation exponents $\eta_{L2}$ and $\eta_{L4}$
in the case $m=2$ of a biaxial {LP}. The consistency of our results
with known expansions about the lower critical dimension $d_\ell=2+m/2$
is demonstrated in section~\ref{sec:lcd}.

In section~\ref{sec:spcases} we first consider the special cases
$(m,d)=(1,4)$ and $(4,5)$. For the former, we derive fully analytical
expressions for the $1/n$ coefficients of the correlation exponents
and the anisotropy exponent. The latter case requires some numerical
work, even though part of the calculations can also be done
analytically. Next, we turn to the physically interesting case
$(m,d)=(1,3)$ of a uniaxial LP in three dimensions and present
numerical results for the coefficients of the $1/n$ contributions. In
section~\ref{sec:uar} we compute the universal scaling function of the
energy-density pair correlation function and a related universal
amplitude to leading order in $1/n$. We give analytical results for
the expansion coefficient of the latter for general values of $(m,d)$
with $d_\ell<d<d^*$. Our results are briefly discussed and put in
perspective in the closing section~\ref{sec:concl}. Finally, there are
two appendixes describing technical details.

\section{Scaling properties of the two-point functions}
\label{sec:scalforms}
We wish to consider the theory of an $n$-component vector field
$\bphi(\bm{x})=\left(\phi_1(\bm{x}),\ldots ,\phi_n(\bm{x})\right)$ in
$d$ space dimensions.  Writing $\bm{x}=(\bm{r},\bm{z})$, we decompose
the position vector $\bm{x}\in\mathbb{R}^d$ into a ``parallel''
component $\bm{z}\in \mathbb{R}^m$ and a ``perpendicular'' one
$\bm{r}\in \mathbb{R}^{\bar{m}}$, where $\bar{m}=d-m$ is the
codimension of $m$.  Likewise,
$\partial_{\bm{r}}\equiv\partial/\partial\bm{r}$ and
$\partial_{\bm{z}}=\partial/\partial{\bm{z}}$ denote the
corresponding components of the gradient operator
$\nabla=(\partial_{\bm{r}},\partial_{\bm{z}})$. The Hamiltonian of the
model we will investigate is given by
\begin{equation}
  \label{eq:Ham}
  {\mathcal H}[\bm{\phi}]={\mathcal H}_0[\bm{\phi}] +{\mathcal H}_{\rm
    int}[\bm{\phi}]\;,
\end{equation}
with the $O(n)$-symmetric free and interacting parts
\begin{equation}
  \label{eq:ham0}
  {\mathcal H}_0[\bm{\phi}]
  =\frac{1}{2}\,{\int}\rmd^dx\left[(\partial_{\bm{r}}\bm{\phi})^2
    +(\partial_{\bm{z}}^2\bm{\phi})^2
    +\mathring{\rho}\,(\partial_{\bm{z}}\bm{\phi})^2
    +\mathring{\tau}\,\phi^2\right]
\end{equation}
and
\begin{equation}
  \label{eq:Hint}
  {\mathcal H}_{\rm int}[\bm{\phi}]=\frac{\lambda}{8}\,
  {\int}\rmd^dx\,\phi^4\;,
\end{equation}
respectively.

In order for the model to have a LP, the dimension $d$ must exceed
$d_\ell(m,n)$, the lower critical dimension, which is believed to be
$d^{O(n)}_\ell(m)=2+m/2$ for the case $n>1$ of continuous $O(n)$
symmetry we are concerned with here (Hornreich \etal 1975a, Grest and
Sak 1978, Diehl 2002).

A cautionary remark should be added here. The arguments of Hornreich
\etal (1975a) and the expansion about the dimension $d=2+m/2$ (Grest
and Sak 1978) actually show only that the \emph{homogeneous ordered}
phase becomes thermodynamically unstable to low-energy excitations at
temperatures $T>0$ whenever $d\leq 2+m/2$.  To establish that a LP
exists for $d>2+m/2$, one would have to prove additionally the
existence of a \emph{modulated ordered} phase that is separated via a
\emph{second-order} line from the disordered phase.  According to an
argument by Garel and Pfeuty (1976) one expects the transition from
the disordered to the modulated ordered phase to be described by a
$2n$-component $\phi^4$ model whose Hamiltonian has $O(n)\times O(n)$
symmetry. Although our results are consistent with the existence of a
LP for $n>1$ and $d>2+m/2$, we cannot rule out the possibility that
for some values of $n$, fluctuations might change this transition into
a discontinuous one. Such a scenario apparently occurs in the ($n=1$)
case of ternary mixtures of A and B homopolymers and AB diblock
copolymers, where fluctuations were found to transform the continuous
mean-field transition between the disordered and lamellar phases into
a discontinuous one (see e.g.\ D{\"u}chs \etal (2003) and its
references), in accordance with general arguments given by
Brazovski\v{\i} (1975).

Whenever a LP exists, we denote the values of $\mathring{\tau}$ and
$\mathring{\rho}$ at which it is located by $\mathring{\tau}_{L}$ and
$\mathring{\rho}_{L}$, respectively.

To derive the large-$n$ expansion, it is convenient to introduce
an auxiliary scalar field $\psi=\psi(\bm{x})$ and rewrite the
interaction part ${\mathcal H}_{\rm int}[\bm{\phi}]$ as
(Br{\'e}zin \etal 1976, Vasiliev \etal 1981a, Vasiliev 1998, Moshe and
Zinn-Justin 2003)
\begin{equation}
\label{eq:Hintref}
\e^{-{\mathcal H}_{\rm int}[\bm{\phi}]}= \mbox{const}\,{\int}{\mathcal
  D}\psi\,
\e^{-\frac{1}{2}{\int}\rmd^dx\,{\left[\psi^2
-i\sqrt{\lambda}\,\phi^2\psi\right]}}\,.
\end{equation}
The full Hamiltonian defined by
equations~(\ref{eq:Ham})--(\ref{eq:Hint}) then becomes
\begin{equation}
\label{eq:fullham}
{\mathcal H}[\bm{\phi},\psi]={\mathcal H}_0[\phi] +\frac{1}{2}\,{\int}
\rmd^dx\,{\left[\psi^2
-\rmi\sqrt{\lambda}\,\phi^2\,\psi\right]},
\end{equation}
up to an irrelevant additive constant.

In view of this reformulation of the model it is natural to consider
correlation functions involving besides the fields $\phi_a$ with
$a=1,\ldots,n$ also $\psi$. We need in particular the two-point
functions. Let us consider the disordered phase. Since in it the
$O(n)$ symmetry of the Hamiltonian~(\ref{eq:fullham}) is not
spontaneously broken, the mixed correlation functions
$\langle\phi_a\,\psi\rangle$ vanish.  Furthermore, invariance under
translations and rotations in the position subspaces ${\mathbb R}^m$
and ${\mathbb R}^{\bar{m}}$ implies
\begin{equation}
  \label{eq:gphi}
  \left\langle\phi_{a_1}(\bm{x}+\bm{x}')\,\phi_{a_2}(\bm{x}')\right\rangle=
  \delta_{a_1a_2}\,G_\phi(r,z)
\end{equation}
and
\begin{equation}
  \label{eq:gpsi}
  \langle\psi(\bm{x}+\bm{x}')\,\psi(\bm{x}')\rangle= G_\psi(r,z),
\end{equation}
where $r\equiv|\bm{r}|$ and $z\equiv|\bm{z}|$ are the lengths of the
perpendicular and parallel components of $\bm{x}=(\bm{r},\bm{z})$.
Writing the wave-vector conjugate to $\bm{x}$ as
$\bm{k}=(\bm{p},\bm{q})$, we introduce Fourier transforms
$\tilde{G}_{\phi,\psi}(p,q)$ of the two-point functions
$G_{\phi,\psi}(r,z)$ via
\begin{equation}
  \label{eq:FTgs}
  G_{\phi,\psi}(r,z)={\int_{\bm{k}}^{(d)}}\tilde{G}_{\phi,\psi}(p,q)\,
    \e^{\rmi\bm{k}\cdot\bm{x}},
\end{equation}
where
\begin{equation}
  \label{eq:kint}
  {\int_{\bm{k}}^{(d)}}\equiv
  \int_{\mathbb{R}^d}\frac{\rmd^dk}{(2\pi)^d} =
  {\int_{\bm{p}}^{(\bar{m})}}{\int_{\bm{q}}^{(m)}}
\end{equation}
is a convenient short-hand.

At a LP, the correlation functions $G_{\phi,\psi}$ are expected to
decay as powers of $r$ and $z$ in the long-distance limits
$r,\,z\to\infty$ provided $d_\ell^{O(n)}<d<d^*$. Furthermore, they
should be anisotropically scale invariant. Thus, for appropriate
choices of the values of the scaling dimensions $\Delta_\phi$ and
$\Delta_\psi$ of the fields $\phi_a$ and $\psi$ and of the anisotropy
exponent $\theta$, the limits
\begin{equation}
  \label{eq:gasphi}
 \lim_{b\to\infty} b^{2\Delta_\phi}\,G_\phi(br,b^{\theta}z)= G^{(\rm
   as)}_\phi(r,z)
\end{equation}
and
\begin{equation}
  \label{eq:gaspsi}
  \lim_{b\to\infty}b^{2\Delta_\psi}\,G_\psi(br,b^{\theta}z)= G^{(\rm
    as)}_\psi(r,z)
\end{equation}
should exist, yielding nontrivial asymptotic functions
$ G^{(\rm  as)}_{\phi,\psi}(r,z)$.
Similar results must hold for their Fourier transforms, namely
\begin{equation}
  \label{eq:gtildeas}
 \lim_{b\to\infty}b^{-2\tilde{\Delta}_{\phi,\psi}}\,
\tilde{G}^{\rm (as)}_{\phi,\psi}(b^{-1}p,b^{-\theta}
 q)= \tilde{G}_{\phi,\psi}^{\rm (as)}(p,q),
\end{equation}
where the scaling dimensions are given by
\begin{equation}
  \label{eq:ys}
  2 \tilde{\Delta}_{\phi,\psi}= d-m/2-2\Delta_{\phi,\psi}.
\end{equation}

The asymptotic functions~(\ref{eq:gtildeas}) are generalized
homogeneous functions satisfying
\begin{equation}
  \label{eq:Gasscf}\fl
\tilde{G}_{\phi,\psi}^{\rm (as)}(p,q)=p^{-2\tilde{\Delta}_{\phi,\psi}}\,
\tilde{G}_{\phi,\psi}^{\rm(as)}(1,p^{-\theta}q)=
q^{-2\tilde{\Delta}_{\phi,\psi}/\theta}\,\tilde{G}_{\phi,\psi}^{\rm
  (as)}\big(q^{-{1/\theta}}p,1\big).
\end{equation}
The exponents governing the momentum dependences of
$\tilde{G}^{\rm (as)}_\phi$ are
conventionally written as
\begin{equation}
\label{eq:etaL2id}
2\tilde{\Delta}_\phi=2-\eta_{L2},\quad
2\tilde{\Delta}_\phi/\theta=4-\eta_{L4},
\end{equation}
which defines the correlation exponents $\eta_{L2}$ and $\eta_{L4}$.

In the limit $n\to \infty$, the momentum-space propagators at the LP
become
\begin{equation}
  \label{eq:gzeros}
  \tilde{G}^{(0)}_\phi(p,q)=\left(p^2+q^4\right)^{-1}, \quad\quad
  \tilde{G}^{(0)}_\psi(p,q)=1\,,
\end{equation}
so that the exponents take their known spherical-model values
\begin{equation}
  \label{eq:spmodval}
  \eta_{L2}^{(0)}=\eta_{L4}^{(0)}=0\;,\quad \theta^{(0)}=1/2\;.
\end{equation}
For finite $n$, fluctuations modify both the propagators and the
critical exponents.  Making the ansatzes
\begin{equation}
 \label{eq:expoexp}
\eta_{L2}(m,d)={\eta_{L2}^{(1)}\over n}+\Or\big(n^{-2}\big)\,,
\quad
\theta(m,d) =\frac{1}{2}+{\theta^{(1)}\over n}+\Or\big(n^{-2}\big),
\end{equation}
we shall determine the coefficients $\eta_{L2}^{(1)}$ and
$\theta^{(1)}$ of the $1/n$ corrections in what follows. For
$\eta_{L4}$ the $1/n$~expansion is analogous to that of $\eta_{L2}$;
the scaling relation~(\ref{eq:etaL2id}) implies that its
$1/n$~coefficient is given by
\begin{equation}
  \label{eq:etafouroneovn}
\eta_{L4}^{(1)}=2\big(\eta_{L2}^{(1)}+4\theta^{(1)}\big)\;.
\end{equation}

\section{Self-consistent equations}
\label{sec:sceq}

The self-consistent equations that must be solved in order to
determine the $\Or(1/n)$ corrections to the correlation exponents can
be derived by well-known steepest descent or diagrammatic methods,
which are amply described in the literature (see e.g.\ Ma 1973, Abe
and Hikami 1973, Br{\'e}zin \etal 1976, Vasiliev \etal 1981a, Vasiliev
1998, Moshe and Zinn-Justin 2003). Solving them is, however, another
story: Owing to the anisotropic character of scale invariance at LPs,
the correlation functions $\tilde{G}_\phi$ and $\tilde{G}_\psi$
involve scaling \emph{functions}, rather than being simple powers.
Finding solutions to these equations is therefore a nontrivial task.

We use an appropriately modified variant of the technique advocated by
Vasiliev \etal (1981a, 1981b) and employed to compute the critical
exponents $\eta$ and $\nu$ of the standard $n$-component
$|\bm{\phi}|^4$ model up to order $n^{-2}$.  The crux of this method
is to make a scaling ansatz for the asymptotic large-length-scale form
of the solutions and determine the exponents it involves from the
conditions implied by the self-consistent equations.

These  equations can be written as
\begin{eqnarray}
  \label{eq:scGphi}
  {G}_\phi^{-1}&=&G_\phi^{(0)-1}-\Sigma_\phi=G_\phi^{(0)-1}-
\;\raisebox{-14pt}{\includegraphics[width=30pt]{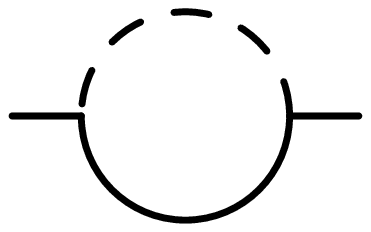}}\; -\ldots,\\
\label{eq:scGpsi}
  {G}_\psi^{-1}&=& G_\psi^{(0)-1}-\Sigma_\psi
  =G_\psi^{(0)-1}-\raisebox{-8.5pt}{\includegraphics[width=35pt]{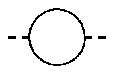}}  -\ldots ,
\end{eqnarray}
where the full and broken lines denote the full
propagators~(\ref{eq:gphi}) and (\ref{eq:gpsi}), respectively.

Since the propagator~(\ref{eq:gphi}) is proportional to
$\delta_{a_1a_2}$, each closed loop of $G_\phi$ lines produces a
factor $n$.  We assume that $G_\phi=\Or(1)$ as $n\to\infty$.  Then it
follows from equation~(\ref{eq:scGpsi}) that $G_\psi^{-1} \sim n$ as
$n\to \infty$.  For the self-energy function $\Sigma_\psi$ an
expansion in inverse powers of $n$ follows, with the leading
contribution being of order $n$. Consequently $G_\psi=\Or(1/n)$, and
since $G_\phi=\Or(1)$, only a finite number of graphs contribute to
$\Sigma_\phi$ and $\Sigma_\psi$ in calculations up to a given order in
$1/n$.  If we restrict ourselves to the first nontrivial order $1/n$,
we may drop all contributions represented by the ellipsis in these
two equations.

According to equations~(\ref{eq:gzeros}), the free parts
$\tilde{G}_\psi^{(0)-1}$ and $\tilde{G}_\phi^{(0)-1}$ are analytic in
the momenta $p$ and $q$. Since the asymptotic forms of the
self-consistent solutions involve nontrivial powers of momenta for
generic $d$, as we shall see, these free parts do not contribute to
$\tilde{G}_{\phi,\psi}^{\rm (as)}(p,q)$ and may be discarded. From
equation~(\ref{eq:scGpsi}) we thus find
 \begin{equation}
  \label{eq:Gpsias}
  \tilde{G}_\psi^{\rm (as)}(p,q)=\frac{2}{n\,\lambda\,F(p,q)}\;,
\end{equation}
while equation~(\ref{eq:scGphi}) yields
\begin{equation}
  \label{eq:Gphias}
\big[\tilde{G}^{({\rm as})}_\phi(p,q)\big]^{-1}=
\frac{2}{n}\int_{\bm{p}'}^{(\bar{m})}\int_{\bm{q}'}^{(m)}
\frac{\tilde{G}^{\rm (as)}_\phi(|\bm{p}'+\bm{p}|,
  |\bm{q}'+\bm{q}|)}{F(p',q')}
\end{equation}
with
\begin{equation}
  \label{eq:bubbleI}
   F(p,q)=\int_{\bm{p}'}^{(\bar{m})}\int_{\bm{q}'}^{(m)}
   \tilde{G}^{\rm (as)}_\phi(|\bm{p}'+\bm{p}|,
   |\bm{q}'+\bm{q}|) \,\tilde{G}^{\rm (as)}_\phi(p^\prime,q^\prime).
\end{equation}
Just as $\tilde{G}_{\phi,\psi}^{\rm (as)}(p,q)$, $F(p,q)$ is a
generalized homogeneous function; equation~(\ref{eq:bubbleI}) can be
combined with the scaling form~(\ref{eq:Gasscf}) of $\tilde{G}^{(\rm
  as)}_\phi$ to conclude that
\begin{eqnarray}
  \label{eq:Sidf1}
F(p,q)&=&p^{\bar{m}+\theta m-4\tilde{\Delta}_\phi}\,F(1,p^{-\theta}q)\\ &=&
q^{(\bar{m}+\theta
  m-4\tilde{\Delta}_\phi)/\theta}\,F(q^{-{1/\theta}}p,1)\,.
\end{eqnarray}

We can now insert these expressions together with their
analogues~(\ref{eq:Gasscf}) for $\tilde{G}_{\phi,\psi}^{\rm
  (as)}(p,q)$ into equation~(\ref{eq:Gphias}), choosing either one of
the components $\bm{p}$ and $\bm{q}$ of the external momentum to be
zero.  Matching the amplitudes of the corresponding powers
$p^{2\tilde{\Delta}_\phi}$ and $q^{2\tilde{\Delta}_\phi/\theta}$ on
both sides of the resulting two equations yields the consistency
conditions
\begin{equation}\label{eq:sctbs}
  \frac{n}{2}=\int^{(\bar{m})}_{\bm{p}}\,
  \frac{p^{4\tilde{\Delta}_\phi-\bar{m}}}{|\bm{p}
    +\bm{1}|^{2\tilde{\Delta}_\phi}} \,
  \int^{(m)}_{\bm{q}}\frac{\tilde{G}^{\rm (as)}_\phi
\big(1,p^\theta|\bm{p} +\bm{1}|^{-\theta}\,q\big)}{F(1,q)}\equiv I_1(n)
\end{equation}
and
\begin{equation}
   \label{eq:scrbs2}\fl
\frac{n}{2}=
\int_{\bm{q}}^{(m)}\frac{q^{4\tilde{\Delta}_\phi/\theta-m}}
{|\bm{q}+\bm{1}|^{{2\tilde{\Delta}_\phi/\theta}} }\,
\int_{\bm{p}}^{(\bar{m})}\frac{\tilde{G}^{\rm
    (as)}_\phi\big({q^{1/\theta} \,
    |\bm{q}+\bm{1}|^{-1/\theta}}\,p,1\big)}{F(p,1)}\equiv I_2(n)\,,
\end{equation}
where $\bm{1}$ designates an arbitrary unit vector in $\bar{m}$ or $m$
dimensions.

Since the left-hand sides are of order $n$, so must be the right-hand
sides. Hence the integrals $I_1(n)$ and $I_2(n)$ must have simple
poles at $1/n=0$. Consider first $I_1(n)$. Its inner integral
$\int^{(m)}_{\bm{q}}$ approaches a $\bm{p}$-independent value as
$p\to\infty$. Recalling equations~(\ref{eq:etaL2id}) and
(\ref{eq:expoexp}), we see that the ratio of $\bm{p}$-dependent
factors in front of this integral behaves as $\sim
p^{2-\bar{m}+\Or(1/n)}$ in this limit.  Had we regularized the
$\bm{p}$-integral by a large-$\bm{p}$ cutoff $\Lambda$, it would have
ultraviolet (uv) divergences of the form $\Lambda^{2-s+\Or(1/n)}$,
$s=0,1,2$. In the dimensionally regularized theory, uv divergences
$\sim\Lambda^{\Or(1/n)}=1+\Or(1/n)\,\ln\Lambda+\ldots$ become poles at
$1/n=0$. Conversely, contributions from the inner integral
$\int^{(m)}_{\bm{q}}$ falling off faster than $p^{-2}$ as $n=\infty$
do not contribute to the residue of the pole $\sim (1/n)^{-1}$, and
hence may be dropped when calculating it.

The upshot is that the $\Or(n)$ contribution of $I_1(n)$ can be
determined as follows. We add and subtract from $\tilde{G}^{(\rm
  as)}_\phi(1,p^\theta|\bm{p} +\bm{1}|^{-\theta}\,q)$ its Taylor
expansion in the second variable $u_p(q)\equiv p^\theta|\bm{p}
+\bm{1}|^{-\theta}\,q$ about its limiting value for $p\to\infty$,
$u_\infty(q)=q$, to the order necessary to ensure that no
contributions of order $n$ are produced by the difference. Details of
this calculation are described in \ref{app1}.  The result is
\begin{equation}
  \label{eq:I1exp}
  I_1(n)=n\,\frac{K_{\bar{m}}}{\eta_{L2}^{(1)}\,\bar{m}}\int_{\bm{q}}^{(m)}
  \frac{{\mathcal P}_1(q^4)}{{(1+q^4)}^3\,I(1,q)}+\Or(n^0)\;.
\end{equation}
Here $K_{\bar{m}}$ is a standard geometric factor defined by
\begin{equation}
  \label{eq:KD}
K_D\equiv (4\pi)^{-D/2}\,\frac{2}{\Gamma(D/2)}\;,
\end{equation}
while ${\mathcal P}_1$ means the polynomial
\begin{equation}
  \label{eq:Pcal1}
  {\mathcal P}_1(q^4)= 4-\bar{m}\,(1+q^4)\,.
\end{equation}
The integral $I(1,q)$, defined by
\begin{equation}
  \label{eq:Ipq}
  I(p,q)=\int_{\bm{p}'}^{(\bar{m})}\int_{\bm{q}'}^{(m)}
  \frac{1}{({p^\prime}^2+{q^\prime}^4)
\left(|\bm{p}'+\bm{p}|^2+|\bm{q}'+\bm{q}|^4\right)}\;,
\end{equation}
is the counterpart of Ma's (1973) critical ``elementary bubble''
$\Pi(k)$ for the present case of $m$-axial {LPs}.

The function $I(p,q)$ satisfies scaling relations similar to those of
$\tilde{G}_{\phi,\psi}^{\rm as}(p,q)$ in equation~(\ref{eq:Gasscf}):
\begin{equation}\label{eq:Scali}
I(p,q)=p^{-\epsilon}\,I\big(1,qp^{-1/2}\big)=q^{-2\epsilon}
\,I\big(pq^{-2},1\big)\,.
\end{equation}
Further properties of this function are discussed in \ref{app2}.

The $\Or(n)$ term of $I_2(n)$ can be worked out in a similar fashion.
To calculate it, we must add and subtract the Taylor expansion of the
corresponding inner integral to fourth order in one of its variables
(see \ref{app1}). Proceeding in this manner yields
\begin{equation}\fl
  \label{eq:I2exp}
  I_2(n)=\frac{n}{2}\,\frac{K_{m}}{\big(\eta_{L2}^{(1)}+4\theta^{(1)}\big)m(m+2)}
\int_{\bm{p}}^{(\bar{m})}
  \frac{{\mathcal P}_2(p^2)}{{(1+p^2)}^5\,I(p,1)}+\Or(n^0)
\end{equation}
with the polynomial
\begin{eqnarray}
  \label{eq:Pcal2}
  {\mathcal P}_2(p^2)&=&3(8-m)(6-m)+5(m^2+2m-96)p^2\\\nonumber
&&\strut +(m^2+50m+144)p^4-m(m+2)p^6\,.
\end{eqnarray}

Using the expressions~(\ref{eq:I1exp}) and (\ref{eq:I2exp}) for the
integrals $I_1(n)$ and $I_2(n)$, we now solve the
equations~(\ref{eq:sctbs}) and (\ref{eq:scrbs2}) for $\eta_{L2}^{(1)}$ and
$\theta^{(1)}$.  This gives
\begin{eqnarray}\label{eq:E2r}
&&\eta_{L2}={1\over n}\;{2K_{\bar{m}}\over \bar{m}}
\int_{\bm{q}}^{(m)}{{\mathcal P}_1(q^4)\over (1+q^4)^3}\;{1\over I(1,q)}
+\Or\big(n^{-2}\big)\,,
\\ \label{eq:Trel}&&
\theta={1\over 2}-\frac{\eta_{L2}}{4}+\frac{1}{n}\,{K_m\over 4m(m+2)}
\int_{\bm{p}}^{(\bar{m})}\!{{\mathcal P}_2(p^2)\over (1+p^2)^5}\,
{1\over I(p,1)}
+\Or\big(n^{-2}\big)\,.
\end{eqnarray}
The implied $1/n$~expansion of $\eta_{L4}$ follows with the aid of
relation~(\ref{eq:etafouroneovn}) for the $1/n$~coefficient
$\eta_{L4}^{(1)}$; it reads
\begin{equation}
 \label{eq:E4r}
\eta_{L4}={1\over n}\;{2K_m\over m(m+2)}\,
\int_{\bm{p}}^{(\bar{m})}{{\mathcal P}_2(p^2)\over (1+p^2)^5}\;{1\over I(p,1)}
+\Or\big(n^{-2}\big)\,.
\end{equation}

In equations~(\ref{eq:Trel}) and (\ref{eq:E4r}) we have expressed the
$1/n$ coefficients in terms of an integral over the perpendicular
momentum $\bm{p}$.  Upon exploiting the scaling
property~(\ref{eq:Scali}) of the integral $I(p,q)$, we could rewrite
this as an integral over the parallel momentum $\bm{q}$. The
transformed integrand would contain the polynomial ${\tilde{\mathcal
    P}}_2(q^4)=q^{12}\,{\mathcal P}_2(q^{-4})$. Likewise, the
$\bm{q}$-integral in equation~(\ref{eq:E2r}) can be recast as an
integral over $\bm{p}$.

\begin{figure}[htb]
\begin{center}
\includegraphics[width=200pt]{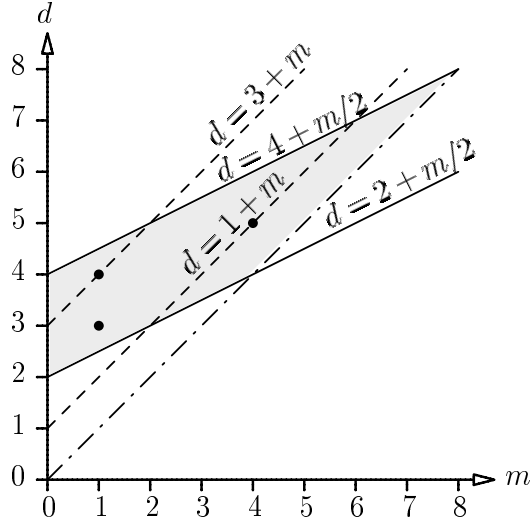}
\end{center}
\caption{Lines of upper and lower critical dimensions $d^*(m)=4+m/2$
  and  $d_\ell(m)=2+m/2$, respectively. In the shaded region, bounded by
  portions of these lines together with the condition $0\leq m\leq d$,
  nonclassical critical behaviour is expected to occur. The full
  circles indicate the special cases discussed in
  section~\ref{sec:spcases}. Also shown are the lines $d=3+m$ and
  $d=1+m$ on which the free propagator simplifies to a simple
  exponential and incomplete gamma function, respectively (see
  section~\ref{sec:spcases}).}%
\label{fig:lines}%
\end{figure}

The above equations~(\ref{eq:E2r}), (\ref{eq:Trel}) and
(\ref{eq:E4r}), giving the expansions of the correlation exponents
$\eta_{L2}$, $\eta_{L4}$ and the anisotropy exponent $\theta$ to order
$1/n$ for general values of $(m,d)$ with $2+m/2<d<4+m/2$, are the main
results of this paper. In the following we further analyse these
results for special choices of $(m,d)$. Figure~\ref{fig:lines}
illustrates the region between the upper and lower critical lines in
which nonclassical behaviour is expected and displays the special
cases to be discussed below as full circles.

\section{The limiting cases of critical points and isotropic Lifshitz
  points}
\label{sec:checkE}

In this section we check that our results for general $m$,
equations~(\ref{eq:E2r})--(\ref{eq:E4r}), comply with known series
expansion to $\Or(1/n)$ for the cases $m=0$ of CPs (Ma 1973, Abe and
Hikami 1973) and $m=d$ of isotropic LPs (Hornreich \etal 1975b). In
either case we must take limits $D\to 0$ of $D$-dimensional integrals
to determine the $\Or(1/n)$ coefficients, where $D=m$ or $\bar{m}$. To
do this, we use the relation
\begin{equation}\label{eq:Dto0}
\lim_{D\to 0}\int\!\!d^Dx\,g(x;D)=\lim_{D\to
  0}S_D\!\int\!\!dx\,x^{D-1} \,g(x;D)
 =g(0;0)\;.
\end{equation}

Consider first the CP case $m\to 0$. Here only the correlation
exponent $\eta_{L2}$ is physically meaningful. From
equation~(\ref{eq:E2r}) one easily derives
  \begin{eqnarray}
\lim_{m\to 0}\eta_{L2}&=&{2\over d}\,K_d\,\lim_{m\to 0}
{{\mathcal P}_1(0)\over I(1,0)}\,{1\over n}+\Or\big(1/n^{2}\big)
\nonumber\\&=&
{2\over d}\,K_d\,{4-d\over J_d(1,1)}\,{1\over n}+\Or\big({1/n^2}\big)\;,
\end{eqnarray}
where $J_d(1,1)$ is a standard one-loop integral, corresponding to a
special case of the quantity $J_D(f,t)$ defined by
equation~(\ref{eq:JDdef}) of \ref{app1}. Its explicit value is given
in equation~(\ref{eq:I2d}). Inserting it into the last equation we
immediately recover the familiar result for $\eta$ first derived by Ma
(1973) and Abe and Hikami (1973):
\begin{equation}
  \label{eq:eta2mzero}
  \eta\equiv\left.\eta_{L2}\right|_{m=0}=\frac{4(4-d)\,
    \Gamma(d-2)}{d\,
    \Gamma(2 - d/2)\,\Gamma^2(d/2-1)\,\Gamma(d/2)}\, \frac{1}{n}
  +\Or\big(1/n^2\big)\;.
\end{equation}

The limit $m\to d$, corresponding to an isotropic LP, can be handled
in a similar fashion. Now only the correlation exponent $\eta_{L4}$ is
physically significant. Equation~(\ref{eq:E4r}) yields
\begin{eqnarray}
 \label{eq:Blid}
\lim_{m\to d}\eta_{L4}&=&
{2K_d\over d(d+2)}\,\lim_{m\to d}{{\mathcal P}_2(0)\over I(0,1)}
\,{1\over n} +\Or\big({1/n^2}\big)\nonumber\\&=&
6\,{(d-8)(d-6)\over d(d+2)}\,{K_d\over J_d(2,2)}\,{1\over n}
+\Or\big({1/n^2}\big)
\end{eqnarray}
with $J_d(2,2)$ given by equation~(\ref{I4d}). Upon making several
transformations of Gamma functions, we recover the result of Hornreich
\etal (1975b) for $\eta_{L4}$, namely
\begin{equation}
\label{eq:eta4md}
\left.\eta_{L4}\right|_{m=d}=3(8-d)\,2^{d-2}\, \frac{\sin(\pi
  d/2)\,\Gamma[(d-3)/2]}{\pi^{3/2}\,d(d+2)\,\Gamma(d/ 2)}\,\frac{1}{n}
+\Or\big({1/n^2}\big)\;.
\end{equation}

\section{Consistency with $\bm{\epsilon}$ expansion about the upper
  critical dimension}
\label{sec:epsexp}

In (Diehl and Shpot 2000, Shpot and Diehl 2001) the $\epsilon$ expansions
about the upper critical dimension $d^*=4+m/2$ of all critical,
crossover, and the usual correction-to-scaling exponents have been
obtained for general $m$ and $n$. These series can be expanded in
powers of $1/n$ to $\Or(1/n)$. Conversely, considering the limit of
small $\epsilon$, our above $\Or(1/n)$ results for general $(m,d)$ can be
expanded in powers of $\epsilon$.  The resulting two double series
expansions in $\epsilon$ and $1/n$ of each exponent should agree.

We shall work out the $\Or(\epsilon^2/n)$ contributions implied by the
$1/n$~expansions (\ref{eq:E2r}) and (\ref{eq:E4r}) for general $m$,
expressing them in terms of multi-dimensional integrals.
Unfortunately, these integrals are in general rather complicated and
not necessarily analytically tractable. For this reason, we will
content ourselves here with verifying the consistency of the series
expansions of $\eta_{L2}$ and $\eta_{L4}$ in $\epsilon$ and $1/n$ for
the special choice $m=2$. Owing to the simple form which the scaling
functions of the position-space propagators take in this case both for
$n=\infty$ and the Gaussian theory (simple exponentials, see
Mergulh{\~a}o and Carneiro (1999), Diehl and Shpot (2000), Shpot and
Diehl (2001)), the $\Or(\epsilon^2/n)$ terms of the series expansions
can be worked out analytically in a straightforward fashion (see
below).

One source of $\epsilon$ dependence in equations (\ref{eq:E2r}) and
(\ref{eq:E4r}) is the function $I(p,q)$. This is just the one-loop
Feynman integral associated with the four-point graph %
\raisebox{-3pt}{\includegraphics[width=30pt]{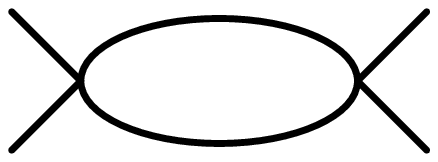}}\, %
of the usual $\phi^4$ theory for $m$-axial {LPs}. Its Laurent
expansion in $\epsilon$ reads
\begin{equation}\label{Sf1}
I(p,q)={p^{-\epsilon}\over\epsilon}\;c_{-1}
\left[1-\epsilon\,F\big({q^2/p}\big)\right]+\Or(\epsilon)\,,
\end{equation}
where $c_{-1}$ and minus the scaling function $F$ represent the
residuum and finite part of $I(1,q)$, respectively.

In the case of $\eta_{L2}$, a second source of $\epsilon$ dependence
is the function ${\mathcal P}_1(q^4)$ (see equation~(\ref{eq:Pcal1})).
Upon decomposing it as
\begin{equation}
  \label{eq:P1dec}
  {\mathcal P}_1(q^4)={\mathcal P}_1^{(0)}(q^4)+
\epsilon\,{\mathcal P}_1^{(1)}(q^4)
\end{equation}
with
\begin{equation}\label{P01}
{\mathcal P}_1^{(0)}(q^4)={m\over 2}\,(1+q^4)-4q^4
\end{equation}
and
\begin{equation}
  \label{eq:P11}
  {\mathcal P}_1^{(1)}(q^4)=1+q^4\;,
\end{equation}
we arrive at the $\epsilon$ expansion
\begin{equation}\fl
\eta_{L2}={1\over n}\;{2K_{\bar{m}}\over \bar{m}}{\epsilon\over c_{-1}}
\int_{\bm{q}}^{(m)}{{\mathcal P}_1^{(0)}(q^4)+
\epsilon\,{\mathcal P}_1^{(1)}(q^4)
\over (1+q^4)^3}\left[1+\epsilon\,F(q^2)\right]
+\Or\Big(\frac{\epsilon^3}{n},\frac{\epsilon^2}{n^2}\Big)\,.
\end{equation}

The result suggests the presence of a term linear in $\epsilon$.
However, the $\epsilon$ expansions of the correlation exponents
$\eta_{L2}$ and $\eta_{L4}$ are known to start at order $\epsilon^2$,
so this term must vanish. It does indeed, since
\begin{equation}\label{Z1}
\int_0^\infty\!\!dq\,q^{m-1}\; {{\mathcal
    P}_1^{(0)}(q^4)\over(1+q^4)^3}=0\;.
\end{equation}
Hence we have
\begin{equation}\label{eq:E2ne}\fl
\eta_{L2}=\frac{\epsilon^2}{n}\,\frac{4K_{4-m/2}}{(8-m)\,c_{-1}}\,
{\int_{\bm{q}}^{(m)}}\frac{{\mathcal P}_1^{(0)}(q^4)\,F(q^2)
  +{\mathcal P}_1^{(1)}(q^4)}{(1+q^4)^3}
+\Or\Big(\frac{\epsilon^3}{n},\frac{\epsilon^2}{n^2}\Big)\;.
\end{equation}

Turning to the expansion~(\ref{eq:E4r}) of $\eta_{L4}$, we note that
the polynomial ${\mathcal P}_2(p^2)$ in its integrand does not depend
on $\epsilon$. However, the integration measure contains a factor
$p^{-\epsilon}$ which cancels the one originating from $I(p,1)$ (see
equation~(\ref{Sf1})). It follows that
\begin{equation}
\label{eq:E4re}\fl
\eta_{L4}=\frac{1}{n}\,\frac{2K_m}{m(m+2)}\frac{\epsilon}{c_{-1}}\,
\int_{\bm{p}}^{(\bar{m}^*)}\frac{{\mathcal P}_2(p^2)}{(1+p^2)^5}\;
\left[1+\epsilon\,F(1/p)\right]
+\Or\Big({\epsilon^3\over n}\,,{\epsilon^2\over n^2}\Big)\,.
\end{equation}
Similarly as above, the contribution from the 1 in the square brackets
vanishes because
\begin{equation}\label{Z2}
{\int_0^\infty}dp\,p^{3-m/2}\frac{{\mathcal P}_2(p^2)}{{(1+p^2)}^5}=0
\end{equation}
for general $m$. Thus the contribution linear in $\epsilon$ is zero,
and the $\epsilon$ expansion of the large-$n$ result~(\ref{eq:E4r})
becomes
\begin{equation}
\label{eq:ZI2}
\eta_{L4}={\epsilon^2\over n}\;{2K_m\over m(m+2)\,c_{-1}}
\int_{\bm{p}}^{(\bar{m}^*)}\frac{{\mathcal P}_2(p^2)\,F(1/p)}{(1+p^2)^5}\;
+\Or\Big({\epsilon^3\over n}\,,{\epsilon^2\over n^2}\Big)\,.
\end{equation}

Equations (\ref{eq:E2ne}) and (\ref{eq:ZI2}) give the contributions of
$\Or(\epsilon^2/n)$ implied by our large-$n$ results for general $m$.
They should agree with those obtained from the $\epsilon$-expansion
results of (Diehl and Shpot 2000, Shpot and Diehl 2001). Let us verify
this explicitly for the biaxial case $m=2$, for which both the residue
$c_{-1}$ as well as the function $F$ appearing in the Laurent
expansion~(\ref{Sf1}) may be gleaned from Mergulh{\~a}o and Carneiro
(1999). Taking into account that the expansion parameter
$\epsilon_\|$, called $\epsilon$
by these authors, corresponds to $\epsilon_\|=2\epsilon$ in our notation,%
\footnote{Note that there is a misprint in the corresponding formula
  of Shpot and Diehl (2001) which precedes its equation (69). The
  correct relation between $\epsilon_\|$ and $\epsilon=d^*(m)-d$ is as
  given above in the main text.}
one sees that
\begin{eqnarray}\label{Fm1}
\left.c_{-1}\right|_{m=2}&=&\frac{1}{32\pi^2}\;,\\
\left.F(q^2)\right|_{m=2}&=&\frac{q^2}{2}\,\arctan\big({2}/{q^2}\big)+
\frac{1}{2}\ln\big(1+{q^4}/{4}\big)-1-\ln4\;.
\end{eqnarray}

Upon substituting this into equations~(\ref{eq:E2ne}) and
(\ref{eq:ZI2}), one can perform the required integrals to obtain
\begin{eqnarray}
\eta_{L2}(m=2)&=&{4\over9}\,{\epsilon^2\over n}+
\Or\Big({\epsilon^3\over n}\,,{\epsilon^2\over n^2}\Big)\;,\\
\eta_{L4}(m=2)&=&-{8\over27}\,{\epsilon^2\over n}
+\Or\Big({\epsilon^3\over n}\,,{\epsilon^2\over n^2}\Big)\,.
\end{eqnarray}
Expanding the correct $\Or(\epsilon^2)$ results (Sak and Grest 1978,
Mergulh{\~a}o and Carneiro 1999, Shpot and Diehl 2001) to order $1/n$
yields identical results.

Since the Laurent expansion of the integral $I(p,q)$ to order
$\epsilon^0$ is also explicitly known when $m=6$ (where $d^*=7$)
(Mergulh{\~a}o and Carneiro 1999), a similar consistency check should
also be possible in this case via analytical, albeit somewhat more
complicated, calculations.

A complete proof of agreement of the results of the $1/n$ and
$\epsilon$ expansions for general $m$ lies beyond the scope of the
present work. The main reason why we have so far been unable to
generalize the foregoing consistency check to general values of $m$ is
our lack of knowledge of a sufficiently simple closed form for the
finite part of $I(p,q)$.

\section{Consistency with expansions about the lower critical dimension}
\label{sec:lcd}

The $1/n$~expansions (\ref{eq:E2r}) and (\ref{eq:E4r}) should hold
down to the lower critical dimensionality $d_\ell$, which is given by
the line $d_\ell(m)=2+m/2$ in our case (unless the LP gets destroyed
by fluctuations). Considering dimensions $d=d_\ell+\epsilon_\ell$
slightly above $d_\ell$, we can additionally expand in
$\epsilon_\ell$. Conversely, any pertinent series expansion in
$\epsilon_\ell$ that is available can be expanded in powers of $1/n$.
The resulting pairs of double power series in $1/n$ and $\epsilon$
must agree.

Series expansion about the lower critical dimension can be obtained
for systems of the kind we are concerned with here---namely, systems
having a low-temperature phase with spontaneously broken $O(n)$
symmetry---by analysing appropriate nonlinear sigma models (Polyakov
1975, Br{\'e}zin and Zinn-Justin 1976a, 1976b, Bardeen \etal 1976).  There
are two sorts of results we can compare with: $\epsilon_\ell$ expansions for
the CP case $m=0$ (Br{\'e}zin and Zinn-Justin 1976b) on the one hand, and
for $m$-axial LPs (Grest and Sak 1978), on the other hand. By
investigating an appropriate generalization of the conventional
nonlinear sigma model, the latter authors produced one-loop results
for the three special values $m=1$, $m=2$, and $m=4$.

To show that our $\Or(1/n)$ results~(\ref{eq:E2r})--(\ref{eq:E4r}) are
compatible with these $\epsilon_\ell$ expansions, we proceed in much
the same way as in section~\ref{sec:epsexp}.  The integrals
$I(p,1)$ and $I(1,q)$ in the denominators of these equations have a
pole at $\epsilon_\ell=0$. According to equations~(\ref{eq:Jst}) and
(\ref{eq:Ist}),
\begin{equation}\label{Ltpole}
I(p,1)=I(1,q)+\Or\big(\epsilon_\ell^0\big)=(4 \pi)^{-1-{m/4}}\,
{\Gamma({m/4})\over \Gamma({m/ 2})}\;
{2\over\epsilon_\ell}+\Or\big(\epsilon_\ell^0\big)\,.
\end{equation}
Substituting this into equations~(\ref{eq:E2r})--(\ref{eq:E4r}), one
can perform the remaining integrals. Unlike the case of small $\epsilon$,
nonzero contributions linear in $\epsilon_\ell$ appear. We obtain
\begin{equation}
\eta_{L2}=\frac{\eta_{L4}}{2}
+\Or\Big({\epsilon_\ell^2\over n},{\epsilon_\ell\over n^2}\Big)=
{\epsilon_\ell\over n}+\Or\Big({\epsilon_\ell^2\over n},
{\epsilon_\ell\over n^2}\Big)
\end{equation}
and
\begin{equation}
  \label{eq:thetaepsl}
  \theta={1\over 2}+\Or\Big({\epsilon_\ell^2\over n},
{\epsilon_\ell\over n^2}\Big)\,.
\end{equation}

These results are consistent  both with Grest and Sak (1978) as well as
with Br{\'e}zin and Zinn-Justin (1976b). Since they were derived for
general $m$, the above critical exponents are \emph{independent} of
$m$ to the indicated order in $1/n$ and $\epsilon$. Note that the
$\Or(\epsilon_\ell/n)$ correction to the anisotropy exponent vanishes.

From the structure of the $\epsilon_\ell$ expansions it is clear that
the $\Or(\epsilon_\ell)$ terms of $\eta_{L2}$ and $\eta_{L4}$ must be
inversely proportional to $n-2$. Hence we can generalize Grest and
Sak's (1978) results for $m=1$, $2$ and $4$ to conclude that
\begin{eqnarray}
&&\eta_{L2}(m)={1\over2}\eta_{L4}(m)+\Or(\epsilon_\ell^2)=
{\epsilon_\ell\over n-2}+\Or(\epsilon_\ell^2)\,,
\\&&
\theta(m)={1\over 2}+\Or(\epsilon_\ell^2)\,,
\end{eqnarray}
for general $m$.

As one sees, the series coefficients of the terms linear in
$\epsilon_\ell$ are independent of $m$. This is a feature they have in
common with the coefficients of the $\Or(\epsilon)$ contributions of
\emph{all} bulk critical exponents of the $m$-axial LPs we are
concerned with here (Shpot and Diehl 2001). For the $\epsilon$
expansion of the above exponents this is trivially fulfilled since
their $\Or(\epsilon)$ terms are zero. However, other exponents
such as those associated with the parallel and perpendicular
correlation lengths do have $\Or(\epsilon)$ contributions, whose
coefficients do not depend on $m$.

\section{Special cases}
\label{sec:spcases}

In this section we further exploit our results for general $m$,
equations (\ref{eq:E2r}), (\ref{eq:Trel}) and (\ref{eq:E4r}), by
considering special cases. Of primary physical interest clearly is the
case $(m,d)=(1,3)$ of a three-dimensional system with uniaxial
anisotropy. Unfortunately, this case appears too difficult to allow a
completely analytic calculation of the $\Or(1/n)$ terms. Let us
therefore first consider some simpler cases, before returning to it.

\subsection{The case $m=1$, $d=4$}

For this choice of $(m,d)$, the required calculations can be done
analytically to obtain closed expression for the $\Or(1/n)$
coefficients of the exponents $\eta_{L2}$, $\eta_{L4}$ and $\theta$.

To see this, recall that the free propagator in position space takes a
simple exponential form on the whole line $d=3+m$ (Mergulh{\~a}o and
Carneiro 1999, Shpot and Diehl 2001). On it, the scaling function
$\Phi_{m,d}(\upsilon)$ of the Fourier back transform
\begin{equation}\label{eq:Sgrz}
G_\phi^{(0)}(r,z)=r^{-2+\epsilon}\,\Phi_{m,d}(v)\,,\quad
\upsilon\equiv z/\sqrt r\;,
\end{equation}
of the free momentum-space propagator $\tilde{G}_\phi^{(0)}(p,q)$
introduced in equation (\ref{eq:gzeros}) simplifies to
\begin{equation}\label{eq:Pgrz}
\Phi_{m,3+m}(\upsilon)=(4\pi)^{-2+\epsilon}\e^{-\upsilon^2/4},
\end{equation}
whilst away from it, it generally is a difference of two generalized
hypergeometric functions, a so-called Fox-Wright $_1\Psi_1$ function
(Shpot and Diehl 2001).

Previously, this simplifying feature was exploited in the context of
the $\epsilon$ expansion in two different ways: Mergulh{\~a}o and
Carneiro (1999) fixed the codimension $\bar{m}=d-m$ at $\bar{m}=3$ and
took $m=2-2\epsilon$ (and hence $d=5-2\epsilon$) to expand about the
point $(m,d)=(2,5)$. Shpot and Diehl (2001) performed a two-loop RG
analysis for general fixed $m$ and dimensions $d=d^*(m)-\epsilon$.
Owing to the simple form~(\ref{eq:Pgrz}), both types of calculations
could be performed analytically for $m=2$, as well as for $m=6$ where
one can benefit from similar simplifications at $d^*(6)=7$. Despite
the difference of the two procedures, they yielded consistent results
for the $\epsilon$~expansions of the critical exponents about
$(m,d)=(2,5)$ to $\Or(\epsilon^2)$.\footnote{Verifying the consistency
  of Mergulh{\~a}o and Carneiro's (1999) series expansions to
  $\Or(\epsilon^2)$ about $(m,d)=(2,5)$ and $(6,7)$ with those of
  Shpot and Diehl (2001) requires the correction of two minor mistakes
  in the former reference; see section 4.2 of Shpot and Diehl (2001)
  for details.}

Since the $\epsilon$~expansion is asymptotic, it gives us direct
information only about the behaviour in the immediate vicinity of the
line of upper critical dimensions $d^*(m)$. In order to derive from it
reliable predictions for the values of critical exponents in $d=3$
dimensions, one must extrapolate, for example, in the case $m=1$ of a
uniaxial LP down to $\epsilon=3/2$. This is not normally possible
unless sophisticated extrapolation and resummation methods are
employed. The $1/n$~expansion, on the other hand, does not require
$\epsilon$ to be small and hence enables us to move further away from
$d^*$. Here we exploit the simple form~(\ref{eq:Pgrz}) to gain
information about the behaviour at $(m,d)=(1,4)$, corresponding to a
distance of $\epsilon=1/2$ from $d^*(1)=9/2$.

It is no complicated matter to compute the function $I(1,q)$ for $m=1$
and $d=4$; one obtains
\begin{equation}
  \label{Ixfin}
I(1,q)\big|_{m=1}^{d=4}
={1\over 8\pi\sqrt 2}\sqrt{\sqrt{q^4+4}-q^2}
={1\over 4\pi\sqrt 2}{1\over\sqrt{\sqrt{q^4+4}+q^2}}\,,
\end{equation}
which yields $I(p,1)=p^{-1/2}\,I(1,p^{-1/2})$ by virtue of the scaling
property~(\ref{eq:Scali}). Upon inserting these expressions into
equations~(\ref{eq:E2r})--(\ref{eq:E4r}), one can perform the required
integrals in a straightforward manner. This gives the following
results for the $1/n$~coefficients of the exponents $\eta_{L2}$, $\theta$
and $\eta_{L4}$:
\begin{equation}
  \label{E14}
\left.
  \begin{array}[c]{l}
\eta_{L2}^{(1)}={5\over 9\pi\sqrt 3}\simeq 0.1021\;,\\[\medskipamount]
\theta^{(1)}=-{4\over 27\pi\sqrt 3}\simeq -0.0272\,,\\[\medskipamount]
\eta_{L4}^{(1)}=-{2\over 27\pi\sqrt 3}\simeq  -0.0136\,,
  \end{array}\right\}\;\; m=1\,,\;d=4\,.
\end{equation}

Despite the smallness of the $\Or(1/n)$ corrections, these results
provide clear evidence for the fact that the critical exponents
$\eta_{L2}$, $\eta_{L4}$ and $\theta$ have nonclassical values below
the upper critical dimension. This conclusion is in full accord with
previous work based on the $\epsilon$~expansion (Diehl and Shpot 2000,
Shpot and Diehl 2001).

\subsection{The special case $m=4$, $d=5$}

There is another line on which the scaling function $\Phi_{m,d}$
simplifies: For $d=m+1$ it reduces to an incomplete gamma function
(see equation~(18) of Shpot and Diehl (2001)). Specifically for
$d=m+1=5$, it becomes the elementary function
\begin{equation}
  \label{eq:Pphi45}
\Phi_{4,5}(\upsilon)={1\over 2(2\pi)^2}\,{1\over \upsilon^2}
\left(1-\e^{-\upsilon^2/4}\right)\,.
\end{equation}

%The point $(m,d)=(4,5)$ is also special in that it lies on the line
%$\epsilon=1$ (i.e.\ $d=3+m/2$) where the derivative $\Pi'(r)$ of the ``elementary
%bubble'' $\Pi(r)$, considered by Ma (1973), Suzuki (1973), and also by
%Inayat-Hussain and Buckingham (1990) in other contexts, has a
%logarithmic singularity in the asymptotic regime $r\to 0$ when
%$0<m<d$.

Its simplicity enables us to compute the integral $I(1,q)$ for this
choice of $m$ and $d$ without much difficulty. We obtain
\begin{equation}\fl
I(1,q)\big|_{m=4}^{d=5}
=\frac{1}{2(2\pi)^2}\,{1\over 4q^2}\bigg\{
q^2\arctan\bigg[\frac{2}{q^2(q^4+3)}\bigg]+\ln{1+q^4\over 1+ q^4/4}\bigg\}
\end{equation}
and a corresponding result for $I(p,1)=p^{-1}I(1,p^{-1/2})$.  These
expressions can be substituted into
equations~(\ref{eq:E2r})--(\ref{eq:E4r}) and the required integrals
then evaluated by numerical means. This yields the values
\begin{equation}
\left.
  \begin{array}[c]{l}
\eta_{L2}^{(1)}\simeq 0.314\;,\\[\medskipamount]
\theta^{(1)}\simeq -0.0728\,,\\[\medskipamount]
\eta_{L4}^{(1)}\simeq 0.045\,,
  \end{array}\right\}\;\; m=4\,,\;d=5\,,
\end{equation}
for the $1/n$ coefficients of $\eta_{L2}$, $\theta$ and $\eta_{L4}$.
Remarkably, the $1/n$ coefficient of $\eta_{L4}$ is no longer
negative, as it was both for infinitesimally small $\epsilon$ and at
$(m,d)=(1,4)$.

\subsection{The case $m=1$, $d=3$}

We now turn to the uniaxial, three-dimensional case $(m,d)=(1,3)$.
Unfortunately, we have not been able to evaluate fully analytically
the required integrals of the $1/n$~coefficients.

Let us start from equation~(\ref{eq:Ipq}). The
($\bar{m}=2$)-dimensional integration over the perpendicular momentum
$\bm{p}'$ can be performed in a straightforward fashion, giving
\begin{eqnarray}\label{eq:Ilog}\fl
I(1,q)
 ={1\over 8\pi^2}\int_{-\infty}^\infty dq'
A^{-1/2}\ln\biggl[{2q'^4(q'+q)^4+A+\big[q'^4+(q'+q)^4+1\big]A^{1/2} \over
  2q'^4(q'+q)^4}\biggr],
\end{eqnarray}
where  $A$ stands for the expression
\begin{equation}\label{eq:Gqy}
A(q',q)=\left[1+(q'+q)^4\right]^2+(1+q'^4)^2-2q'^4(q'+q)^4-1\,.
\end{equation}

The integration over $q'$ can be regarded as an integral in
the complex $q'$ plane and after some work be shown to reduce to%
\footnote{We are indebted to S.\ Rutkevich for this calculation.}
\begin{equation}\label{eq:Serg}
I(1,q)=-{i\over 2\pi}\int_0^\frac{i}{2q}{dq'\over
\sqrt{(1+4q'^2q^2)\left[1+\case14 (4q'^2+q^2)^2\right]}}.
\end{equation}
This in turn can be expressed as a complete elliptic integral or
Gauss hypergeometric function to obtain
\begin{equation}\label{Ielf}
I(1,q)=\case{1}{4}{(4+q^4)}^{-1/4}{(1+q^4)}^{-1/2}
\,_2F_1\big({\case{1}{2}},\case{1}{2};1;k^2\big)
\end{equation}
with
\begin{equation}
k^2={1\over2}\bigg[1-
{3+q^4\over (1+q^4){(1+4/q^4)}^{1/2}}\bigg].
\end{equation}

Known quadratic and linear transformation formulae for the
hypergeometric function $_2F_1$, such as equations~(15.30.30) and
(15.3.4) of Abramowitz and Stegun (1972), enable us to cast the above
integral in the following two equivalent forms
\begin{eqnarray}\label{Ielf1}
I(1,q)&=&{1\over4\sqrt
  2}\,u^{1/4}\,_2F_1\big(\case{1}{4},\case{1}{4};1;u\big)\,,
\quad
u={4\over(1+q^4)^2(4+q^4)}\;,\\
&=&{1\over4\sqrt
  2}\,w^{1/4}\,_2F_1\big(\case{1}{4},\case{3}{4};1;-w\big)\,,
\quad
w={4\over q^4(3+q^4)^2}\,.
\end{eqnarray}
Upon substitution of either one of them into our general expressions
(\ref{eq:E2r})--(\ref{eq:E4r}) for the $\Or(1/n)$ coefficients (along
with their counterparts for $I(p,1)=p^{-3/2}I(1,p^{-1/2})$), the
remaining integral over $q$ can easily be performed numerically. The
results are
\begin{eqnarray}\label{eq:E231}
\eta_{L2}(m=1,d=3)&\simeq& 0.306\,{1\over n}+\Or(n^{-2})\,,
\\[\medskipamount]
\theta(m=1,d=3)&\simeq&{1\over2} -0.0487\,{1\over n}+\Or(n^{-2})\,,
\\[\medskipamount] \label{E431}
\eta_{L4}(m=1,d=3)&\simeq& 0.223\,{1\over n}+\Or(n^{-2})\,.
\end{eqnarray}

Just as in the previous case of $m=4$, $d=5$, and unlike the
coefficient of the $\Or(\epsilon^2/n)$ term, $\eta_{L4}^{(1)}$ is
\emph{positive}. This suggests a tendency of $\eta_{L4}$ to change
from small negative values of $d\lesssim d^*$ to positive ones as $d$ is
lowered further.

\section{A universal scaling function and related amplitude} \label{sec:uar}

So far we have focused our attention on the calculation of critical
exponents. However, the $1/n$~expansion can also be employed to
compute other universal quantities such as universal amplitude ratios
and scaling functions. As an example, we here compute the leading
nontrivial term of a universal scaling function associated with the
energy-density cumulant at the LP and a related amplitude.

According to general scaling arguments and RG work (Diehl and Shpot
2000, Shpot and Diehl 2001, Diehl \etal 2003a, 2003b), the leading singular
part of this function is expected to take on large length-scales the
asymptotic scaling form
\begin{equation}
  \label{eq:Gphi2}
  \tilde{G}_{\phi^2}^{\rm (sing)}(p,q)
\equiv\left.\langle\phi^2\phi^2\rangle^{\rm cum}_{p,q}\right|_{\rm
  sing} \approx
  E_1\,p^{-\alpha_L/\nu_{L2}}\,\Psi_{m,d}(E_2\,qp^{-\theta})\;.
\end{equation}
Here the quantity on the left-hand side is the Fourier transform of
the cumulant $\langle\phi^2(\bm{x})\phi^2(\bm{0})\rangle^{\rm cum}$,
where ``sing'' means singular part. Further, $\alpha_L$ and $\nu_{L2}$
are the usual critical indices of the specific-heat and perpendicular
correlation length; they are related to the exponents used
above via
\begin{equation}
  \label{eq:alphaL}
   \alpha_L/\nu_{L2}= 2/\nu_{L2}-\bar{m}-m\theta=2\tilde{\Delta}_\psi\;.
\end{equation}
Furthermore, $E_1$ and $E_2$ denote nonuniversal metric factors. The
former can be varied by changing the normalization of the
order-parameter field $\phi$, the latter by modifying the scale in
which parallel momenta are measured, i.e.\ by multiplying the term
$(\partial_z^2\bphi)^2$ of the Hamiltonian by a factor $\sigma\neq 1$.

The scaling function $\Psi_{m,d}$ should be universal, but depends of
course on $n$. We choose its normalization such that
$\Psi_{m,d}(0)=1$. To fix the scale of its argument, we could require
that the logarithmic derivative of $\Psi_{m,d}({\sf q})$ at a
reference value ${\sf q}_0$ (for instance, ${\sf q}_0=0$) takes a
certain value. We do this simply by making the choice $E_2=1$. The
$1/n$~expansion of $\Psi_{m,d}$ then starts at order $(1/n)^0$.
Taking into account that the function $\tilde{G}_{\phi^2}(p,q)$ is
trivially related to $\tilde{G}_\psi(p,q)$ (see e.g.\ chapter 29 of
Zinn-Justin (1996)) and recalling equation~(\ref{eq:scGpsi}), one sees
that
\begin{equation}
  \label{eq:Psimd}
  \Psi_{m,d}({\sf q})=\frac{I(1,{\sf q})}{I(1,0)}+\Or(1/n)\;.
\end{equation}
If we take the limit $p\to 0$ at fixed $q$, the scaling
form~(\ref{eq:Gphi2}) must yield a $p$-independent contribution $\sim
q^{-\alpha_L/\nu_{L2}\theta}$. Hence the scaling function must have the
asymptotic behaviour
\begin{equation}
  \label{eq:Psimdinft}
  \Psi_{m,d}({\sf q}\to \infty)\approx {\mathcal C}(m,d)\,{\sf
    q}^{-\alpha_L/(\nu_{L2}\theta)}\;.
\end{equation}

From equation~(\ref{eq:Psimd}) we conclude that the universal
${\mathcal C}(m,d)$ coefficient, to leading order in $1/n$, is given
by
\begin{eqnarray}
{\mathcal C}(m,d)={\mathcal C}^{(0)}(m,d)+\Or(1/n) \;\;\mbox{ with
}\;\; {\mathcal C}^{(0)}(m,d) =\frac{I(0,1)}{I(1,0)}\,.
\end{eqnarray}
In \ref{app2} we show that the required two integrals can both be
calculated explicitly for general values of $m$ and $d$. The results
can be found in equations~(\ref{eq:Jst}) and (\ref{eq:Ist}). We
refrain from giving the resulting rather lengthy expression for
${\mathcal C}(m,d)$ here, restricting ourselves to a discussion of
some special cases of interest.

\begin{figure}[t]
\centerline{\includegraphics[width=300pt]{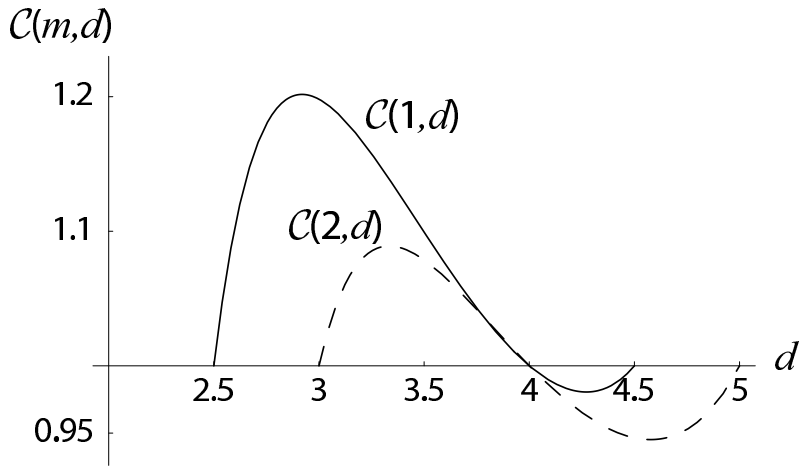}}%[width=11cm,height=7cm]
\caption{Universal amplitude ${\mathcal C}(m,d)$ as a function of $d$
  for $m=1$ (full line) and $m=2$ (broken line)}%
\label{fig:Cmds}%
\end{figure}

Consider, first, the uniaxial case $m=1$. Here our result becomes
\begin{equation}
\label{eq:C1d}\fl
{\mathcal C}^{(0)}(1,d)=\frac{2^{- d+3/2}\,
      {\sqrt{\pi }}\,
      \Gamma\left[ (2d-3)/4\right]\,
      \sin \left[( 2\,d-5)
            \,\pi/4)\right]\,
      \left[ 2^d +
        16\,\sin (d\,\pi/2)
        \right] }{\cos (d\,\pi )\Gamma(
       {1}/{4})\,
      \Gamma[(d-1)/2]}\;.
\end{equation}
As shown in figure~\ref{fig:Cmds}, this coefficient is a smooth and
finite function of $d$ in the whole range $d_\ell=5/2\leq d\leq d^*=9/2$.
It has the $\epsilon$~expansion
\begin{equation}
{\mathcal C}^{(0)}(1,{9}/{2}-\epsilon\big) =1+(\pi-4 -2\ln
2) \,\frac{\epsilon}{12}
+\Or(\epsilon^2)\,.
\end{equation}

The corresponding results for the biaxial case $m=2$ are
\begin{equation}\fl
{\mathcal C}^{(0)}(2,d)=\pi^{-3/2}\cot(d\pi/2)\left[
\Gamma[(5-d)/2]\,\Gamma[(d-4)/2]-4^{5-d}\pi\frac{\Gamma(d-4)}{\Gamma(d-7/2)}
\right]
\end{equation}
and
\begin{equation}
{\mathcal C}^{(0)}(2,5-\epsilon)=1+(\ln 2-1)\,\epsilon
+\Or(\epsilon^2)\,.
\end{equation}
Again, this coefficient behaves smoothly between the upper and lower
critical dimensions $d_\ell=3$ and $d^*=5$ (see figure~\ref{fig:Cmds}).

\section{Summary and Discussion} \label{sec:concl}

In this paper we have shown how to utilize the $1/n$~expansion for the
study of critical behaviour at $m$-axial {LPs}. We have been able to
determine the $\Or(1/n)$ corrections of the correlation exponents
$\eta_{L2}$ and $\eta_{L4}$, and of the related anisotropy exponent
$\theta$, for general values of $m$ and $d$ with $0\leq m\leq d$ and
$2+m/2<d< 4+m/2$. What makes such calculations a challenge is a
combination of two problems: the anisotropic scale invariance one
encounters already at the level of the free theory, and the
complicated forms of the propagators' scaling functions at the LP.

To cope with these difficulties, it proved useful to employ a properly
adjusted and generalized modification of the technique Vasiliev \etal
(1981a, 1981b) introduced to compute the critical exponents $\eta$ and
$\nu$ of the standard $n$-component $|\bm{\phi}|^4$ model up to order
$n^{-2}$. We believe that the present work may serve as a starting
point for more ambitious studies based on the $1/n$~expansion. One
question which might be systematically investigated in this manner is
whether the predictions for scaling functions of anisotropic scale
invariant systems made recently by Henkel (1997, 2002) have any
significance in cases where fluctuation effects must not be ignored.

Apart from yielding insights into the feasibility of the approach to
such anisotropic scale invariant system, our work permits us
to draw two interesting conclusions. First of all, it provides
unequivocal evidence for the fact that the critical exponents
$\eta_{L2}$, $\eta_{L4}$ and $\theta$ differ for $d<d^*$ from their
classical values. Although this is in complete accord with what the
available dimensionality expansions about the upper and lower critical
dimensions tell us (see sections~\ref{sec:epsexp} and \ref{sec:lcd}),
it goes considerably beyond these results because no extrapolation in
$d$ is involved in the $1/n$~expansion. This means in particular that
the anisotropy exponent $\theta$ of the three-dimensional uniaxial LPs
should be  nonclassical.

\begin{figure}[t]
  \begin{center}
\includegraphics[width=300pt]{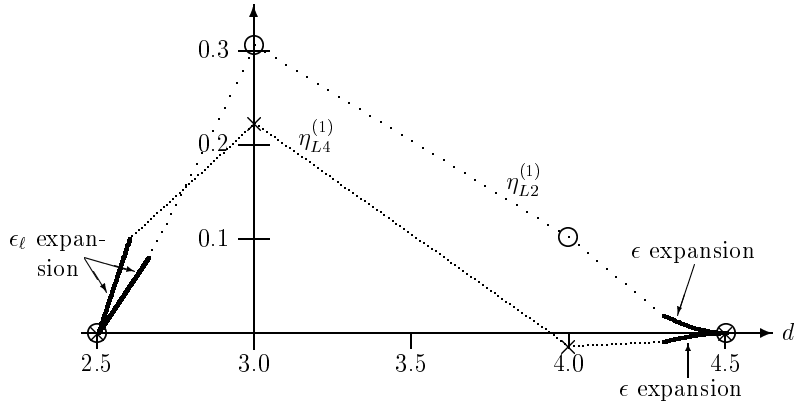}
\caption{Behaviour of the $1/n$~coefficients $\eta_{L2}^{(1)}(1,d)$
  (open circles) and $\eta_{L4}^{(1)}(1,d)$ (crosses) as functions of
  $d$. The 4 points displayed for either one of them correspond to
  results described in section~\ref{sec:spcases}.  The thick lines
  near the upper and lower critical dimensions represent the limiting
  forms $\sim \epsilon^2$ and $\sim \epsilon_\ell$, which the
  dimensionality expansions mentioned in sections~\ref{sec:epsexp} and
  \ref{sec:lcd} yield. The widely or densely dotted lines serve to
  guide the eyes.  The exponent $\eta_{L4}^{(1)}(1,d)$ has a small
  negative value beneath the upper critical dimension $d^*(1)=9/2$ and
  appears to change sign somewhat below $d=4$.}%
\label{fig:etaones}%
  \end{center}
\end{figure}

Another remarkable aspect of our results is the interesting variation
of the $\Or(1/n)$ coefficient of the exponent $\eta_{L4}$ with $d$. As
illustrated in figure~\ref{fig:etaones} for the uniaxial case $m=1$,
it first decreases to small negative values as $d$ drops below $d^*$,
then becomes positive as $d$ is lowered further, before it drops back
to zero at the lower critical dimension $d_\ell$. If we accept the
plausible hypothesis that the correlation exponents are continuous
functions of $d$, then it seems reasonable to expect a
qualitatively similar behaviour of the exponent $\eta_{L4}$. In other
words, $\eta_{L4}$ should change sign somewhere below $d^*$.  In fact,
there are other examples where the lowest nontrivial term of the
dimensionality expansion about $d^*$ is negative, whereas
extrapolations of higher-order calculations yield a positive value in
$d=3$ dimensions. This happens for instance in the case of the Ising
model with quenched random bond disorder. Here the dimensionality
expansion of the correlation exponent $\eta$ about $d=4$ begins with a
negative contribution $\sim (\sqrt{4-d})^2$ (see e.g.\ Jayaprakash and
Katz 1977), whereas the extrapolation of higher-order calculations
yields a positive value for this exponent at $d=3$ (Pelissetto and
Vicari 2000).

Let us close with a few cautionary remarks. Our results to $\Or(1/n)$
are mathematically well defined in the whole range where $2+m/2<d<
4+m/2$ and $0\leq m \leq d$ (the shaded region in
figure~\ref{fig:lines}).  However, applying them to real
three-dimensional systems requires some care.  They should be combined
with information from other sources such as dimensionality expansions,
simulations, exact results and experiments, keeping in mind their
limitations.  For example, depending on whether $m=2$ or $m>2$, the
lower critical dimension $d_\ell$ is $d_\ell=3$ or $d_\ell>3$,
respectively, whenever $n\geq 2$.  Hence, unless the $O(n)$ symmetry
is explicitly broken, no $m$-axial LPs are expected to occur at
nonzero temperatures in $d=3$, which leaves us with the case of
uniaxial LPs in three dimensions. On the other hand, contributions to
the Hamiltonian that break the $O(n)$ symmetry are generically expected,
for example, for magnetic crystals. An analysis of such spin
anisotropies can be found in (Hornreich 1979).

For $m\geq 2$, one must also worry about another type of anisotropies:
space anisotropies breaking the isotropy in the $m$-dimensional
subspace. These give rise to corresponding anisotropic terms quadratic
in $(\partial^2\phi/\partial z_{\alpha}\partial z_{\alpha'})$, which
we also have not taken into account here. Their effects have been
investigated recently within the framework of the $\epsilon$~expansion
(Diehl \etal 2003b).

Last, but not least, let us mention that the technique of matching
asymptotic scaling forms, on which our above analysis was based, can
be extended to determine the $\Or(1/n)$ corrections of the thermal
exponents $\nu_{L2}$, $\nu_{L4}$, $\gamma_L$, and $\alpha_L$ for
$m$-axial LPs (Shpot \etal 2005).

\ack

We are indebted to S~B Rutkevich for deriving the
result~(\ref{eq:Serg}) for the integral~(\ref{eq:Ilog}). One of us
(MASh) thanks P~A Hlushak and V~B Solovyan for their help with Fortran
programming in the numerical calculation involving this integral at a time
when the simplified expression was not yet known.

We gratefully acknowledge support by the Deutsche
Forschungsgemeinschaft (DFG)---in its initial phase via the Leibniz
program (grant no Di 378/2) and Sonderforschungsbereich 237, in its
final phase via DFG grant no  Di 378/3. One of us (YuMP) also enjoyed
support by the Russian Foundation for Basic Research (grant no
03-01-00837) and the Nordic Grant for Network Cooperation with the
Baltic Countries and Northwest Russia FIN-20/2003.  YuMP and MASh thank
Fachbereich Physik at the Universit{\"a}t Essen (now Universit{\"a}t
Duisburg-Essen) for their hospitality.

\appendix

\section{Large-$\bm{n}$ behaviour of momentum integrals}\label{app1}

In subsequent calculations we shall need the properties of the
following $D$-dimensional momentum integral:
\begin{equation}\label{eq:JDdef}
J_D(f,t)=\int_{\bm{k}}^{(D)}\,
k^{-2f}\,|\bm{k}+\bm{1}|^{-2t}={\mathcal V}(f,t;D-f-t)\,,
\end{equation}
where the function ${\mathcal V}$ is defined by
\begin{equation}\label{4.11}\fl
{\mathcal V}(c_1,c_2;c_3)=(4\pi)^{-{D}/{2}}a(c_1)a(c_2)a(c_3)
\,,\quad\quad a(c_i)\equiv
\frac{\Gamma\left({D/2}-c_i\right)}{\Gamma(c_i)}\,,\;\; i=1,2,3.
\end{equation}

Let us consider the parameters $f$ and $t$ of the integral~(\ref{eq:JDdef}) as regular functions of some small parameter $\kappa$.
Choosing  $l\geq 0$ to be integer, we set
\begin{eqnarray}
f+t={D\over 2}-l+\kappa\,.
\label{4.12}
\end{eqnarray}
Then the integral $J_D(f,t)$ is singular in the region of large $k$
for small values of $\kappa$. Its singular part is given by
\begin{equation}
{\mathcal V}_{\mathrm{sing}}\Big(f,t;{D/2}+l-\kappa\Big)\label{Vsing}
=K_D{(-1)^l\over l!}\;
{\left.(f)_l\,(t)_l\right|_{\kappa=0}\over ({D/2})_l}\,{1\over 2\kappa}\,,
\label{4.13}
\end{equation}
where $K_D$ was defined in equation~(\ref{eq:KD}),
and $(\cdots)_l$ are Pochhammer symbols:
\begin{equation}
(w)_l\equiv {\Gamma(w+l)\over\Gamma(w)}=w(w+1)\cdots(w+l-1)\,,
\quad\quad (w)_0=1\,.
\end{equation}
Specifically for $l=0$, $l=1$, $l=2$ and $\kappa\to 0$, we have
\begin{eqnarray}\label{Vs0}
&&{\mathcal V}_{\mathrm{sing}}
\left(f,t;{D/2}-\kappa\right)={K_D\over 2\kappa}\,,\\&&
{\mathcal V}_{\mathrm{sing}}\left(f,t;{D/ 2}+1-\kappa\right)
=-{K_D\over D\kappa}\;f\,t|_{\kappa=0}\,,
\label{4.14a}\\&&\label{4.14b}
{\mathcal V}_{\mathrm{sing}}\left(f,t;{D/2}+2-\kappa\right)
={K_D\over D(D+2)\kappa}\;f(f+1)\,t(t+1)|_{\kappa=0}\,.
\end{eqnarray}

Two relevant examples of the integral $J_D(f,t)$ are
\begin{equation}\label{eq:I2d}\fl
J_d(1,1)=\int_{\bm{p}}^{(d)}\,{1\over p^2\,(\bm{p}+\bm{1})^2}=
(4\pi)^{-{d}/{2}}\,{\Gamma\left(2-{d/2}\right)\over\Gamma(d-2)}\;
\Gamma^2\!\left({d/2}-1\right)
\end{equation}
and
\begin{equation}\label{I4d}\fl
J_d(2,2)=\int_{\bm{q}}^{(d)}\,{1\over q^4\,(\bm{q}+\bm{1})^4}=
(4\pi)^{-\frac{d}{2}}\,{\Gamma\left(4-{d/ 2}\right)\over\Gamma(d-4)}\;
\Gamma^2\!\left({d\over 2}-2\right).
\end{equation}

Now let us consider the momentum integrals $I_1(n)$ and $I_2(n)$
of equations~(\ref{eq:sctbs}) and (\ref{eq:scrbs2}).
We have to isolate their pole contributions at $n=\infty$.
For further convenience we write them as
\begin{eqnarray}
&&I_1(n)=\int^{(\bar{m})}_{\bm{p}}\,
         \frac{p^{4\tilde{\Delta}_\phi-\bar{m}}}{|\bm{p}
         +\bm{1}|^{2\tilde{\Delta}_\phi}}\,\int^{(m)}_{\bm{q}}
G_1\Big({p^{2\theta}\over|\bm{p}+\bm{1}|^{2\theta}}\;q^2\Big)\,{1\over F(1,q)}\,,
\label{E1}\\&&
I_2(n)=\int_{\bm{q}}^{(m)}\frac{q^{4\tilde{\Delta}_\phi/\theta-m}}{
       |\bm{q}+\bm{1}|^{{2\tilde{\Delta}_\phi/\theta}} }\,
       \int_{\bm{p}}^{(\bar{m})}
G_2\Big({q^{2/\theta}\over|\bm{q}+\bm{1}|^{2/\theta}}\;p^2\Big)\,
{1\over F(p,1)}\,,
\label{E2}
\end{eqnarray}
where  $G_1(Q^2)$ and $G_2(P^2)$ denote the scaling functions
$\tilde{G}^{\rm(as)}_\phi(1,Q)$ and $\tilde{G}^{\rm(as)}_\phi(P,1)$.

Adding and subtracting the asymptotic forms
of the arguments of $G_1$ and $G_2$ for $p\to\infty$ and $q\to\infty$
we write these functions as
\begin{eqnarray}\label{4.9a}\fl
&&G_1\!\left(q^2\,p^{2\theta}|\bm{p}+\bm{1}|^{-2\theta}\right)
\equiv G_1(q^2+q^2\alpha_{\bm{p}})
=\sum_{s=0}^2\frac{1}{s!}\,\frac{d^s G_1(q^2)}{d(q^2)^s}
\;q^{2s}\alpha^s_{\bm{p}}+R_1(p,q),
\\\label{4.9b}\fl
&&G_2\!\left(p^2\,q^{2/\theta}|\bm{q}+\bm{1}|^{-2/\theta}\right)
\equiv G_2(p^2+p^2\beta_{\bm{q}})
=\sum_{s=0}^4\frac{1}{s!}\,\frac{d^s G_2(p^2)}{d(p^2)^s}\;
p^{2s}\beta^s_{\bm{q}}+R_2(p, q).
\end{eqnarray}
The deviations
\begin{equation}\label{AB}
\alpha_{\bm{p}}\equiv p^{2\theta}|\bm{p}+\bm{1}|^{-2\theta}-1
\quad\quad\mbox{and}\quad\quad
\beta_{\bm{q}}\equiv q^{2/\theta}|\bm{q}+\bm{1}|^{-{2/ \theta}}-1
\end{equation}
are of order $p^{-1}$ and $q^{-1}$ for large $p$ and $q$.
Including their successive powers enhances the
ultraviolet convergence of $p$ and $q$
integrations in equations~(\ref{E1}) and (\ref{E2}).
The remainders $R_1(p,q)$ and $R_2(p,q)$ of the Taylor expansions
contain sufficiently high powers of $\alpha_{\bm{p}}$ and $\beta_{\bm{q}}$
to make the integrations finite as $n\to\infty$.
These convergent contributions would be required for a
calculation to $\Or(n^{-2})$. All we need now is
\begin{eqnarray}
&&I_1(n)=\sum_{s=0}^2\frac{J^\alpha _s}{s!}\;\lim_{n\to\infty}
\int^{(m)}_{\bm{q}}\,q^{2s}\,
\frac{d^s G_1(q^2)}{d(q^2)^s} {1\over F(1,q)}+\Or(n^0),
\label{4.220a}\\
&&I_2(n)=\sum_{s=0}^4 \frac{J^\beta_s}{s!}\;
\lim_{n\to\infty}\int^{(\bar{m})}_{\bm{p}}\,p^{2s}\;
\frac{d^s G_2(p^2)}{d(p^2)^s}\; {1\over F(p,1)}+\Or(n^0)\,.
\label{4.220b}
\end{eqnarray}
The coefficients $J_s^\alpha$ and $J_s^\beta$ are given by the integrals
\begin{eqnarray}\label{JA}
&&J_s^\alpha=\int^{(\bar{m})}_{\bm{p}}\,p^{4\tilde{\Delta}_\phi-\bar{m}}\,
|\bm{p}+\bm{1}|^{-2\tilde{\Delta}_\phi}\,\alpha^s_{\bm{p}}\,,
\quad\quad s=0,1,2\,,\\\label{JB}
&&J_s^\beta=\int^{(m)}_{\bm{q}}\,
q^{{4\tilde{\Delta}_\phi\over\theta}-m}\,|\bm{q}+\bm{1}|^
{-{2\tilde{\Delta}_\phi\over\theta}}\,\beta^s_{\bm{q}}\,,
\quad\quad s=0,\dots,4\,,
\end{eqnarray}
which have simple poles in $1/n$ at $n=\infty$.  This fact allows us
to take the limit $n\to\infty$ in the remaining integrals over
$\bm{q}$ and $\bm{p}$ in equations~(\ref{4.220a}) and (\ref{4.220b}).

Making binomial expansions of the integer powers $\alpha^s_{\bm{p}}$
and $\beta^s_{\bm{q}}$ in equations~(\ref{JA}) and (\ref{JB}), we can
express $J_s^\alpha$ and $J_s^\beta$ in terms of the
integral~(\ref{eq:JDdef}):
\begin{eqnarray}\label{JAi}\fl
J_s^\alpha=\sum_{j=0}^s(-1)^{s-j}C_s^j\;
{\mathcal V}\!\left({\bar{m}\over 2}
-2\tilde{\Delta}_\phi-j\theta,\;\tilde{\Delta}_\phi+j\theta;
\;{\bar{m}\over 2}+1-{\eta_{L2}^{(1)}\over 2n}\right),
\\\label{JBi}\fl
J_s^\beta=\sum_{j=0}^s(-1)^{s-j}C_s^j\;
{\mathcal V}\!\left[{1\over\theta}\,
\Big({m\theta\over 2}-2\tilde{\Delta}_\phi -j\Big),
{\tilde{\Delta}_\phi+j\over\theta};
{m\over 2}+2-{\eta_{L2}^{(1)}+4\theta^{(1)}\over n}\right]\,,
\end{eqnarray}
where $C_s^j$ are the corresponding binomial coefficients.

Comparing the arguments of the function ${\mathcal V}$ in
equations~(\ref{JAi}) and (\ref{JBi}) with their counterparts in
equation~(\ref{Vsing}), we read off $l=1$ and
$\kappa=\eta_{L2}^{(1)}/(2n)$ for $J_s^\alpha$, and $l=2$ with
$\kappa=(\eta_{L2}^{(1)}+4\theta^{(1)})/n$ for $J_s^\beta$.  With the aid of
the results~(\ref{4.14a}) and (\ref{4.14b}) we then obtain the
required leading large-$n$ behaviour of $J_s^\alpha$ and $J_s^\beta$:
\begin{eqnarray}
&&J_s^\alpha=\frac{K_{\bar{m}}}{\bar{m}}\;
{2n\over\eta_{L2}^{(1)}}\;A_s+\Or(n^0)\,,
\label{4.20a}\\\label{4.20b}
&&J_s^\beta=\frac{K_m}{m(m+2)}\;
{n\over\eta_{L2}^{(1)}+4\theta^{(1)}}\;B_s+\Or(n^0)\,.
\end{eqnarray}
Here  $A_s$ (with $s=0,1,2$) and $B_s$ (with $s=0,\dots,4$)
denote the sums
\begin{eqnarray}\fl
A_s=\sum_{j=0}^s (-1)^{s-j+1}\,C_s^j\,
\Big({\bar{m}\over 2}-2-{j\over 2}\Big)\Big(1+{j\over 2}\Big)\,,
\\\fl
B_s=\sum_{j=0}^s (-1)^{s-j}\,C_s^j\,
\Big({m\over 2}-4-2j\Big)\Big({m\over 2}-3-2j\Big)(2+2j)(3+2j)\,.
\end{eqnarray}
A simple calculation gives:
\begin{eqnarray}\fl
A_0={1\over2}(4-d+m),\;\quad\quad\quad
A_1={1\over 4}(7-d+m),\;\;\quad\quad
A_2=\frac{1}{2}\,;
\nonumber\\\label{4.21}\fl
B_0= \frac{3}{2}(8-m)(6-m),\;\;\quad
B_1={1\over2}(16-m)(66-7m)\,,
\\\fl
B_2= 2(612 - 58 m + m^2),\;\;
B_3= 48 (24-m),\;\;\quad\quad\quad
B_4= 384\,.
\nonumber\end{eqnarray}
In conjunction with the relations~(\ref{4.20a}) and (\ref{4.20b}),
these results give us the coefficients $J_s^\alpha$ and $J_s^\beta$ in
equations~(\ref{4.220a}) and (\ref{4.220b}).

The limit $n\to\infty$ in these equations reduces the scaling
functions in the integrands to those of the free theory.  That is, we
may use the large-$n$ result
\begin{equation}\fl
\lim_{n\to\infty}F(p,q)=I(p,q)=\int_{\bm{k}'}^{(d)}
\frac{1}{({p^\prime}^2+{q^\prime}^4)
\left(|\bm{p}'+\bm{p}|^2+|\bm{q}'+\bm{q}|^4\right)}
\end{equation}
for both scaling functions $F(1,q)$ and $F(p,1)$ introduced in
equation~(\ref{eq:bubbleI}).

Similarly, the scaling functions $G_1(q^2)$ and $G_2(p^2)$
in equations~(\ref{4.220a}) and (\ref{4.220b}) reduce to
\begin{equation}\label{SP}
\tilde{G}_1^{(0)}(q^2)={1\over 1+q^4}\quad\quad\mbox{and}\quad\quad
\tilde{G}_2^{(0)}(p^2)={1\over p^2+1}
\end{equation}
in this limit.

We can now evaluate the sums in equations~(\ref{4.220a}) and
(\ref{4.220b}).  Apart from overall factors, we have
\begin{eqnarray}
&&\sum_{s=0}^2 \frac{A_s}{s!}\;q^{2s}\;\frac{d^s G_1^{(0)}(q^2)}{d(q^2)^s}
=\frac{{\mathcal P}_1(q^4)}{2(1+q^4)^3}\,,
\label{4.24a}\\\label{4.24b}
&&\sum_{s=0}^4 \frac{B_s}{s!}\;p^{2s}\;\frac{d^s G_2^{(0)}(p^2)}{d(p^2)^s}
=\frac{{\mathcal P}_2(p^2)}{2(1+p^2)^5}\,,
\end{eqnarray}
where the polynomials ${\mathcal P}_1(q^4)$ and ${\mathcal P}_2(p^2)$
are given explicitly in equations~(\ref{eq:Pcal1}) and
(\ref{eq:Pcal2}) of the main text.  Finally, equations~(\ref{4.220a})
and (\ref{4.220b}), along with (\ref{4.20a})--(\ref{4.20b}) and
(\ref{4.24a})--(\ref{4.24b}), yield the leading-order expressions of
$I_1(n)$ and $I_2(n)$ given in equations~(\ref{eq:I1exp}) and
(\ref{eq:I2exp}).

\section {The one-loop integral}\label{app2}

The one-loop Feynman integral $I(p,q)$ is an important ingredient of
our calculations.  Unfortunately, we did not succeed to evaluate it
for generic dimensions and external momenta.  Various simplified
expressions which result for special values of $m$ and $d$ are
presented in the main text.  Below, we consider the simplifying cases
of zero external momenta where closed analytic expressions are
obtained for arbitrary $m$ and $d$.

\subsection{The integral $I(p,0)$}\label{Acc}
Consider the integral
\begin{equation}
I(p,0)={\int_{\bm{p}\prime}^{(\bar{m})}}{\int_{\bm{q}\prime}^{(m)}}
\frac{1}{p^{\prime 2}+q^{\prime 4}}\,
{1\over(\bm{p}^{\prime}+\bm{p})^2+q^{\prime 4}}
\end{equation}
It is convenient to convert this integral to
the coordinate representation where it is given
by a Fourier transform of the squared free propagator:
\begin{equation}
I(p,0)=\int\rmd^{\bar m}r\int\rmd^m z
\left[G_\phi^{(0)}(r,z)\right]^2\; \e^{-i\bm{p}\bm{r}}\,.
\end{equation}
Using the scaling representation~(\ref{eq:Sgrz})
for $G_\phi^{(0)}(r,z)$ and changing
the integration variable $z$ via $z=v\sqrt r$ we obtain
\begin{equation}\label{split}
I(p,0)=\int\rmd^{\bar m}r\;r^{-4+{m/2}+2\epsilon}\,
\e^{-i\bm{p}\bm{r}}\int\rmd^m\upsilon\, \Phi^2(\upsilon)\;.
\end{equation}
The $r$ integration is standard, giving
\begin{equation}
\int\rmd^{\bar m}r\; r^{-4+{m/2}+2\epsilon} \; \e^{-i\bm{p}\bm{r}}=
2^\epsilon \,\pi^{\bar{m}/ 2}{\Gamma({\epsilon/ 2})
\over \Gamma(2-{m/ 4}-\epsilon)}\; p^{-\epsilon}\,.
\end{equation}
To calculate the integral over $\upsilon$, we use the Fourier
representation (Diehl and Shpot 2000, Shpot and Diehl 2001)
\begin{equation}
\Phi(\upsilon)=(2\pi)^{-{\bar{m}/ 2}}\int_{\bm{q}}^{(m)}\,q^{\bar{m}-2}\,
K_{\bar m/2-1}(q^2)\,\e^{i\bm{q}\bm{\upsilon}}\,.
\end{equation}
This leads us to
\begin{equation}\label{square}
\int\rmd^m\upsilon\; \Phi^2(\upsilon)=(2 \pi)^{-\bar{m}}\int_{\bm{q}}^{(m)}\;
q^{2\bar{m}-4} K^2_{\bar m/2-1}(q^2)\,.
\end{equation}
The last integral is known for arbitrary $m$ and $d$. We obtain
\begin{equation}\fl
\int\rmd^m\upsilon\; \Phi^2(\upsilon)=(2 \pi)^{-d}\,
\pi^{m/ 2}\;{1\over 2}\; 2^{-\epsilon}\,
\Gamma^2(1-{\epsilon/ 2})\,
{\Gamma(2-{m/ 4}-\epsilon)\over\Gamma(2-\epsilon)}
\;{\Gamma({m/ 4})\over \Gamma({m/ 2})}\,.
\end{equation}
Multiplying both contributions yields
\begin{equation}\label{eq:Jst}
I(p,0)=(4 \pi)^{-{d/ 2}}\,{1\over 2}\,
\Gamma({\epsilon/ 2})\,
{\Gamma^2(1-{\epsilon/ 2})\over\Gamma(2-\epsilon)}\;
{\Gamma({m/ 4})\over \Gamma({m/ 2})}\; p^{-\epsilon}\,.
\end{equation}

\subsection {The integral $I(0,q)$}\label{Aar}

Let us consider another special case of the one-loop integral $I(p,q)$,
\begin{equation}\label{i0e}
I(0,1)=\int_{\bm{p}}^{(\bar{m})}\int_{\bm{q}}^{(m)}
{1\over p^2+q^4}\,{1\over p^2+(\bm{q}+\bm{1})^4}\,.
\end{equation}
Its treatment is somewhat more involved.
We shall use some tricks employed by Mergulh{\~a}o and Carneiro (1999)
in a similar calculation.
Making a partial fraction expansion
\begin{equation}
{1\over (a+b)(a+c)}={1\over c-b}\left({1\over a+b}-{1\over a+c}\right)\;,
\end{equation}
we represent (\ref{i0e}) as
\begin{equation}\fl
I(0,1)=\int_{\bm{q}}^{(m)}{1\over (\bm{q}+\bm{1})^4-q^4}\;
\int_{\bm{p}}^{(\bar{m})}
\left[ {1\over p^2+q^4}-{1\over p^2+(\bm{q}+\bm{1})^4} \right]\,.
\end{equation}
Integrating over $p$ gives
\begin{equation}\label{Trick}\fl
I(0,1)=(4\pi)^{-{\bar{m}/ 2}}\Gamma(-\nu)\int_{\bm{q}}^{(m)}
{1\over (\bm{q}+\bm{1})^4-q^4}\left( q^{4\nu}-|\bm{q}+\bm{1}|^{4\nu}\right)
\end{equation}
with
\begin{equation}
\nu={\bar{m}\over 2} -1=1-{m\over 4}-{\epsilon\over 2}\,.
\end{equation}
It is useful to represent the difference in the last brackets
via the elementary integral
\begin{equation}
a^\alpha-b^\alpha=-\alpha\,(b-a)\int\limits_0^1
{d\,x\over [a+x(b-a)]^{1-\alpha}}
\end{equation}
with $a=q^2$, $b=(\bm{q}+\bm{1})^2$, and $\alpha=2\nu$. The factor
$(b-a)$ cancels with a corresponding term of the denominator (see
equation~(\ref{Trick})). We obtain
\begin{eqnarray}\fl
I(0,1)=\frac{(-2\nu)\,\Gamma(-\nu)}{(4\pi)^{{\bar{m}/2}}}\,
\int_{\bm{q}}^{(m)}{1\over q^2+(\bm{q}+\bm{1})^2}
\int\limits_0^1d\,x{1\over \left( q^2+2\bm{q}\bm{1} x+x \right)^{1-2\nu}}\,.
\end{eqnarray}
Here, we represent the momentum-dependent factors through Laplace
integrals
\begin{equation}\label{Lapl}
a^{-\alpha}={1\over \Gamma(\alpha)}\int\limits_0^\infty d\,x\;
x^{\alpha-1}\e^{-a\,x}\,.
\end{equation}
This yields
\begin{eqnarray}\nonumber\fl
I(0,1)={(-2\nu)\,\Gamma(-\nu)\over (4\pi)^{{\bar{m}/ 2}}\,\Gamma(1-2\nu)}
\\\times
\int\limits_0^1 dx\int\limits_0^\infty dy\int\limits_0^\infty dz\,
z^{-2\nu}\int_{\bm{q}}^{(m)}\e^{-(2q^2+2\bm{q}\bm{1}+1)y}\,
\e^{-(q^2+2\bm{q}\bm{1} x+x)z}\,.
\end{eqnarray}
Performing the Gaussian integration over $q$ we get
\begin{eqnarray}\fl
I(0,1)={(-2\nu)\,\Gamma(-\nu)\over(4\pi)^{{d/2}}\,\Gamma(1-2\nu)}
\int\limits_0^1dx\int\limits_0^\infty dy\int\limits_0^\infty dz\,
{z^{-2\nu}\over (2y+z)^{m/2}}\,\e^{-{y^2-x^2z^2+xz^2+yz\over 2y+z}}\,.
\end{eqnarray}
The integrals over $t$ and $z$ can be decoupled by a rescaling $y=zt$
of the integration variable $y$. This gives
\begin{eqnarray}\fl
I(0,1)={(-2\nu)\,\Gamma(-\nu)\over(4\pi)^{-{d/2}}\,\Gamma(1-2\nu)}
\int\limits_0^1 dx\int\limits_0^\infty {dt\over (2t+1)^{m/2}}
\int\limits_0^\infty dz\,z^{-1+\epsilon}\,
\e^{-z{t^2+t+x-x^2\over 2t+1}}\,.
\end{eqnarray}
The $z$ integral is of the form~(\ref{Lapl}). We obtain
\begin{eqnarray}\fl
I(0,1)={(-2\nu)\,\Gamma(-\nu)\,\Gamma(\epsilon)\over(4\pi)^{d/ 2}\,\Gamma(1-2\nu)}\,
\int\limits_0^1 dx\int\limits_0^\infty{dt\over (2t+1)^{{m/2}-\epsilon}}\,
{(t^2+t+x-x^2)}^{-\epsilon}\,.
\end{eqnarray}
Inside of the last brackets we add and subtract
$1/4$. Next, we introduce the new integration variables
$y=2t+1$ and $s=2x-1$. A little bit of algebra then yields
\begin{eqnarray}\fl
I(0,1)={(-2\nu)\,\Gamma(-\nu)\,\Gamma(\epsilon)
\over(4\pi)^{d/2}\,\Gamma(1-2\nu)}\,{2^{2\epsilon}\over 4}\,
\int\limits_{-1}^1 ds \int\limits_1^\infty dy\;
y^{-{m/2}+\epsilon}(y^2-s^2)^{-\epsilon}\,.
\end{eqnarray}
Here the last integration produces a Gauss hypergeometric function:
\begin{eqnarray}\fl
I(0,1)={(-2\nu)\,\Gamma(-\nu)\,\Gamma(\epsilon)
\over(4\pi)^{{d/2}}\,\Gamma(1-2\nu)}
\;{2^{2\epsilon-1}\over m-2+2 \epsilon}
\int\limits_{-1}^1 ds\,
_2F_1\!\left(\epsilon,\case{m-2+2\epsilon}{4};
\case{2+m+2\epsilon}{4};s^2\right)\,.
\end{eqnarray}
Performing the remaining integral we  finally obtain
\begin{eqnarray}\nonumber
I(0,q)&=&q^{-2\epsilon}\,(4\pi)^{-{d/2}}
{\Gamma(-\nu)\over\Gamma(1-2\nu)}\,
\Gamma(\epsilon)\Gamma(1-\epsilon)\,2^{-2+2\epsilon}
\\[\medskipamount]\label{eq:Ist}
&&\times\left[{\sqrt{\pi}\over\Gamma({3/2}-\epsilon)}
+{4\over 2-m-2\epsilon}\,{\Gamma[(2+m+2\epsilon)/4]\over
\Gamma[(2+m-2\epsilon)/4]} \right]\;,
\end{eqnarray}
where we reintroduced the initially suppressed trivial dependence
on $q$, taking into account that $\nu=1-m/4-\epsilon/2$.

\section*{References}
\begin{harvard}
\item%[Abe73]{Abe73}
Abe R 1973 Expansion of a critical exponent in inverse powers of
  spin dimensionality {\em Progr. Theor. Phys.} {\bf 49} 113

\item%[AH73]{AH73}
Abe R and Hikami S 1973 Critical exponents and scaling relations in $1/n$
  expansion {\em Progr. Theor. Phys.} {\bf 49} 442

\item%[AS72]{AS}
Abramowitz M and Stegun I~A 1972 {\em Handbook of Mathematical Functions with Formulas,
                  Graphs, and Mathematical Tables\/} Applied Mathematics
 Series (National Bureau of Standards: Washington, D C)

\item%[AGM{\etalchar{+}}00]{Aha00}
Aharony O, Gubser S~S, Maldacena J, Ooguri H and Oz Y 2000 Large $N$ field
  theories, string theory and gravity {\em Phys. Rep.} {\bf 323} 183

\item%[AT00]{AT00}
Arnold P and Tom{\'a}{\v{s}}ik B 2000 ${T}_c$ for dilute {B}ose gases:
  Beyond leading order in $1/{N}$ {\em Phys.\ Rev.\ A} {\bf 62} 063604

\item%[BLS76]{BLS76}
Bardeen W~A, Lee B~W and Shrock R~E 1976 Phase transition in the nonlinear
  $\sigma$ model in a $2+\epsilon$ continuum {\em Phys.\ Rev.\ D} {\bf 14} 985

\item%[BZJ00]{BBZ-J00}
Baym G,  Blaizot J-P and Zinn-Justin J 2000 The transition temperature of the
  dilute interacting {B}ose gas for $n$ internal states {\em Europhys.
  Lett.} {\bf 49} 150

\item Bervillier C 2004 Exact renormalization group
  equation for the Lifshitz critical point \emph{Phys. Lett.}
  {\bf 331A} 110

\item Bray A 2002 Theory of phase ordering kinetics \emph{Adv. Phys. [UK]}
  {\bf 51} 481

\item%[Bra75]{Bra75}
Brazovski\v{\i} S~A 1975 Phase transition of an isotropic system to a
  nonuniform state {\em Sov. Phys. JETP} {\bf 41} 85

\item%[Br{\'e}93]{Lne93}
Br{\'e}zin E 1993, ed {\em The large {N}-expansion in quantum field theory and
  statistical physics: from spin systems to 2-dimensional gravity\/}
  (Singapore: World Scientific)

\item%[BGZJ76]{BLZ76}
Br{\'e}zin E, Le Guillou J~C and Zinn-Justin J 1976 Field theoretical
approach to critical phenomena {\em Phase Transitions and Critical
  Phenomena\/} vol.~6, ed  Domb C and Green M~S (London: Academic
Press) chap.~3, pp.~125--247

\item%[BZJ76a]{BZ-J76}
Br{\'e}zin E and Zinn-Justin J 1976a Renormalization of the nonlinear
  $\sigma$ model in $2+\epsilon$ dimensions---application to the {H}eisenberg
  ferromagnets {\em Phys.\ Rev.\ Lett.} {\bf 36} 691

\item%[BZJ76b]{BZ76}
Br{\'e}zin E and Zinn-Justin J 1976b Spontaneous breakdown of continuous symmetries near two dimensions {\em Phys.\ Rev.\ B} {\bf 14} 3110

\item%[CPRV98]{CPRV98}
Campostrini M, Pelissetto A, Rossi P and Vicari E 1998 Two-point
  correlation function of three-dimensional ${O}({N})$ models: {T}he critical
  limit and anisotropy {\em Phys. Rev. E} {\bf 57} 184

\item%[dAL01]{AL01}
de~Albuquerque L~C and Leite M~M 2001 Anisotropic {L}ifshitz point at
  ${O}(\epsilon_{L}^{2})$ {\em J. Phys. A} {\bf 34} L327

\item[] %[Die02]{Die02}
Diehl H~W  2002 Critical behavior at $m$-axial {L}ifshitz points
{\em  Proc.\ of the 5th International Conference Renormalization Group 2002, Tatranska Strba, High Tatra Mountains (Slovakia, March 10--16)}
{\em Acta physica slovaca} {\bf 52} 271 ({\it Preprint\/} cond-mat/0205284)

\item%[Die04]{Die04}
Diehl H~W 2004 Bulk and boundary critical behavior at Lifshitz
points {\em  Proceedings of the 22nd International Conference on Statistical Physics (STATPHYS 22) of the International Union of
  Pure and Applied Physics (IUPAP), Bangalore (India, July 4--9 2004)}
  {\em Pramana---Journal of Physics, to appear\/}  ({\it Preprint\/} cond-mat/0407352)

\item %[DS00]{DS00a}
Diehl H~W and Shpot M  2000 Critical behavior at ${m}$-axial
  {L}ifshitz points: Field-theory analysis and ${\epsilon}$-expansion
  results {\em Phys.\ Rev.\ B} {\bf 62} 12 338 ({\it Preprint\/} cond-mat/0006007)

\item%[DS01]{DS01}
Diehl H~W and Shpot M  2001 Lifshitz-point critical behaviour to
  ${O}(\epsilon^2)$ {\em J. Phys. A} {\bf 34} 9101 ( {\it Preprint\/} cond-mat/0106502)

\item%[DS03]{DS03}
Diehl H~W and Shpot M 2003 Comment on ``Renormalization-group
  picture of the Lifshitz critical behavior'' {\em
  Phys. Rev. B} {\bf 68} 066401 ({\it Preprint\/} cond-mat/0305131)

\item
Diehl H W, Gerwinski A and Rutkevich S 2003a Boundary critical
behavior at $m$-axial Lifshitz points for a boundary plane parallel to the
                  modulation axes \emph{Phys.\ Rev.\ B} {\bf 68}
                  224428 ({\it Preprint\/} cond-mat/0308483)

\item%[DSZ03]{DSZ03}
Diehl H~W, Shpot M~A and Zia R~K~P 2003b Relevance of space anisotropy in
  the critical behavior of $m$-axial Lifshitz points {\em Phys. Rev. B}
  {\bf 68} 224415 ({\it Preprint\/} cond-mat/0307355)

\item%[DMKB94]{DMKB94}
Doherty J~P, Moore M~A, Kim J~M and Bray A~J 1994 Generalizations of the
  Kardar-Parisi-Zhang equation {\em Phys. Rev. Lett.} {\bf 72} 2041

\item%[DG76]{DG76}
Domb C and Green M~S 1976, eds  {\em Phase Transitions and Critical
  Phenomena\/} vol.~6 (London: Academic Press)

\item
D{\"u}chs D, Ganesan G, Fredrickson G H and Schmid F 2003 Fluctuation
effects in ternary AB+A+B polymeric emulsions  {\em Macromolecules} {\bf 36} 9237

\item%[Fis83]{Fis83}
Fisher M~E 1983 Scaling, universality and renormalization group theory
  {\em Critical Phenomena}, ed F~J~W Hahne {\em Lecture Notes in Physics}, vol. 186
(Berlin: Springer-Verlag) pp. 1--139

\item%[F93]{FH93}
Frachebourg L and Henkel M 1993 Exact correlation function at the
  {L}ifshitz points of the spherical model {\em Physica A} {\bf 195} 577

\item%[FTV02]{FTV02}
Franz M, Te{\v{s}}anovi{\'c} Z and Vafek O 2002 QED$_3$ theory of
  pairing pseudogap in cuprates: From d-wave superconductor to
  antiferromagnet via an algebraic Fermi liquid {\em
  Phys. Rev. B} {\bf 66} 054535

\item
Garel T  and Pfeuty P 1976 Commensurability effects on the critical behaviour
                  of systems with helical ordering
\emph{J. Phys. C} {\bf 9} L246

\item%[Gra02a]{Gra02o}
Gracey J~A 2002a Critical exponent $\omega$ at $O(1/N)$ in
  $O(N)\times O(M)$ spin models {\em Nucl. Phys. B} {\bf 644} 433

\item%[Gra02b]{Gra02}
Gracey J~A 2002b Critical exponents in $O(N)\times O(M)$ spin models at
  $O(1/N^2)$ {\em Phys. Rev. B} {\bf 66} 134402

\item%[GS78]{GS78}
Grest G~S and Sak J 1978 Low-temperature renormalization group for the
  {L}ifshitz point {\em Phys. Rev. B} {\bf 17} 3607

\item%[GZJ98]{GZJ98}
Guida R and Zinn-Justin J 1998 Critical exponents of the {$N$}-vector
  model {\em J. Phys. A} {\bf 31} 8103

\item%
Halperin B~I, Hohenberg P~C and Ma S-K 1972 Calculation of dynamic
critical properties using Wilson's expansion methods \emph{Phys.
Rev. Lett.} {\bf 29} 1548

\item Henkel M 1997 Local scale invariance and strongly anisotropic
  equilibrium critical systems \emph {Phys. Rev. Lett.} {\bf 78} 1940

\item Henkel M 2002 Phenomenology of local scale invariance: from
  conformal invariance to dynamical scaling \emph{Nucl. Phys. B}  {\bf
    641} 405

\item Hornreich R~M 1979 Renormalization-group analysis of critical
  modes of a ferromagnet \emph{Phys. Rev. B} {\bf 19} 5914

\item%[Hor80]{Hor80}
Hornreich R~M 1980 The {L}ifshitz point: {P}hase diagrams and critical
  behavior {\em J. Magn. Magn. Mat.} {\bf 15--18} 387

\item%[HB78]{HB78}
Hornreich R~M and Bruce A~D 1978 Behaviour of the critical wavevector near
  a {L}ifshitz point {\em J. Phys. A} {\bf 11} 595

\item%[HLS75a]{HLS75}
Hornreich R~M, Luban M and Shtrikman S 1975a Critical behavior at the onset
  of $\vec{k}$-space instability on the $\lambda$ line {\em Phys. Rev.
  Lett.\/} {\bf 35} 1678

\item%[HLS75b]{HLS75n}
Hornreich R~M, Luban M and Shtrikman S 1975b Critical exponents at a
  {L}ifshitz point to ${O}(1/n)$ {\em Phys. Lett.} {\bf 55A} 269

\item%[HLS77]{HLS77}
Hornreich R~M, Luban M and Shtrikman S 1977 Exactly solvable model exhibiting a multicritical point {\em  Physica A} {\bf 86A} 465

\item%[IHB90]{IHB90}
Inayat-Hussain A~A and Buckingham M~J 1990 Continuously varying critical
  exponents to ${O}(1/n)$ {\em Phys. Rev. A} {\bf 41} 5394

\item%[JK77]{Jkatz77}
Jayaprakash C and Katz H~J 1977 Higher-order corrections to the
  $\epsilon^{1/2}$ expansion of the critical behavior of the random {I}sing
  system {\em Phys.\ Rev.\ B} {\bf 16} 3987

\item%[KO76]{KO76}
Kalok L and Obermair G~M 1976 Competing intraction and critical properties
  in the spherical model {\em J.\ Phys.\ C} {\bf 9} 819

\item%[Lei03]{Lei03}
Leite M~M 2003 Renormalization-group picture of the {L}ifshitz critical
  behavior {\em Phys. Rev. B\/} {\bf 67} 104415

\item%[LW03]{LDW03}
Le Doussal P and Wiese K~J 2003 Functional renormalization group at
  large $N$ for disordered elastic systems, and relation to replica symmetry
  breaking {\em Phys.\ Rev.\ B\/} {\bf 68} 174202

\item%[LW04]{LDW04}
Le Doussal P and Wiese K~J 2004 Derivation of the functional
renormalization group $\beta$-function at order $1/{N}$ for
manifolds pinned by disorder {\em Nucl.\ Phys.\ B\/} {\bf 701} 409

\item%[Ma73]{Ma73}
Ma S-K 1973 Critical exponents above ${T}_c$ to ${O}(1/n)$ {\em Phys.\
  Rev.\ A} {\bf 7} 2172

\item%[MO95]{MO95}
McAvity D~M and Osborn H 1995 Conformal field theories near a boundary in
  general dimensions {\em Nucl.\ Phys.\ B\/} {\bf 455} 522

\item%[MC99]{MC99}
Mergulh{\~a}o C Jr and Carneiro C~E~I 1999 Field-theoretic calculation of
  critical exponents for the {L}ifshitz point {\em Phys. Rev. B\/} {\bf 59} 13954

\item%[MZJ03]{MZ03}
Moshe M and Zinn-Justin J 2003 Quantum field theory in the large $N$
  limit: a review {\em Phys. Rep.\/} {\bf 385} 69

\item%[Muk77]{Muk77}
Mukamel D 1977 Critical behaviour associated with helical order near a
  {L}ifshitz point {\em J. Phys. A\/} {\bf 10} L249

\item%[ML78]{ML78}
Mukamel D and Luban M 1978 Critical behavior at a {L}ifshitz point:
  {C}alculation of a universal amplitude ratio {\em Phys. Rev. B\/} {\bf 18} 3631

\item%[NTCS76]{NTCS76}
  Nicoll J~F, Tuthill G~F, Chang T~S and Stanley H~E 1976
  Renormalization group calculation for critical points of higher
  order with general propagator {\em Phys. Lett. A\/} {\bf 58} 1

\item%[Par80]{Par80}
Parisi G 1980 Field-theoretic approach to second-order phase transitions
  in two- and three-dimensional systems {\em J. Stat. Phys.\/} {\bf 23} 49

\item%[PRV01]{PRV01}
Pelissetto A, Rossi P and Vicari E 2001 Large-$n$ critical behavior of
  ${O}(n)\times{O}(m)$ spin models {\em Nucl. Phys. B\/} {\bf 607} 605

\item%[PV00]{PV00}
Pelissetto A and Vicari E 2000 Randomly dilute spin models: A six-loop
  field-theoretic study {\em Phys.\ Rev.\ B\/} {\bf 62} 6393

\item%[PV02]{PV02}
Pelissetto A and Vicari 2002 Critical phenomena and renormalization
group theory {\em Phys. Rep.} {\bf 368} 549

\item%[Pol75]{Pol75}
Polyakov A~M 1975 Interaction of {G}oldstone particles in two dimensions:
  Applications to ferromagnets and {Y}ang-{M}ills fields {\em Phys. Lett.
  B\/} {\bf 59} 79

\item%[PHA91]{PHA91}
  Privman V, Hohenberg P~C and Aharony A 1991 Universal Critical-Point
  Amplitude Relations {\em Phase Transitions and Critical Phenomena\/}
  vol.~14, ed Domb C and Lebowitz J~L (London: Academic Press)
  pp~1--134

\item%[SG78]{SG78}
Sak J and Grest G~S 1978 Critical exponents for the {L}ifshitz point:
  epsilon expansion {\em Phys. Rev. B} {\bf 17} 3602

\item%[Sel92]{Sel92}
Selke W 1992 Spatially modulated structures in systems with competing
  interactions {\em Phase Transitions and Critical Phenomena\/} vol.~15, ed C Domb and J~L
Lebowitz (London: Academic Press) pp.~1--72

\item%[SD01]{SD01}
Shpot M and Diehl H~W 2001 Two-loop renormalization-group analysis of
  critical behavior at $m$-axial Lifshitz points {\em Nucl. Phys. B} {\bf 612}
  340 ({\em Preprint\/} cond-mat/0106105)

\item%[SPD05]{SPD05}
Shpot M~A, Pis'mak Y~M and Diehl H~W 2005 {\em to be published}

\item%[Vas98]{Vas98}
Vasiliev A~N 1998 {\em Quantum Field Renormalization Group in the Theory of
  Critical Behavior and Stochastic Dynamics\/} (St-Petersburg: PINF Publ.) 1st
  ed. (In Russian). English translation: {\em The field
  theoretic renormalization group in critical behavior theory and stochastic
  dynamics\/} (Chapman and Hall / CRC, 2004)

\item%[VPK81a]{VPH81a}
Vasiliev A~N, Pis{'}mak Y~M and Khonkonen Y~R 1981a Simple method of
 calculating critical indices in the $1/n$
  expansion {\em Teor. Mat. Fiz.} {\bf 46} 157

\item%[VPK81b]{VPH81b}
Vasiliev A~N, Pis{'}mak Y~M and Khonkonen Y~R 1981b $1/n$-expansion:
 calculation of $\eta$ and $\nu$ in order $1/n^2$ for arbitrary dimension
  {\em Teor. Mat. Fiz.} {\bf 47} 291

\item
Vasiliev A~N, Pis{'}mak Y~M and Khonkonen Y~R 1981c $1/n$-expansion:
 calculation of $\eta$ in the order $1/n^3$ by conformal bootstrap
 method
  {\em Teor. Mat. Fiz.} {\bf 50} 127

\item%[WF72]{WF72}
Wilson K~G and Fisher M~E 1972 Critical exponents in 3.99 dimensions
  {\em Phys. Rev. Lett.} {\bf 28} 240

\item%[WK74]{WK74}
Wilson K~G and Kogut J 1974 The renormalization group and the $\epsilon$
  expansion {\em Phys. Reports} {\bf 12C} 75

\item%[ZJ89]{ZJ89}
Zinn-Justin J 1996 {\em Quantum Field Theory and Critical Phenomena\/} (Oxford:
  Clarendon Press) pp. 618--621 International series of monographs on physics
  3rd ed.

\end{harvard}

\end{document}